\documentclass[conference,compsoc]{IEEEtran}

\ifCLASSOPTIONcompsoc
  \usepackage[nocompress]{cite}
\else
  \usepackage{cite}
\fi
\PassOptionsToPackage{table,dvipsnames}{xcolor}   % xcolor 只加载一次
\PassOptionsToPackage{pdftex}{graphicx}           % graphicx 只加载一次
\usepackage{graphicx}          % 已在第1步声明 pdftex 选项
\usepackage{booktabs}
\usepackage[table,dvipsnames]{xcolor}  % 只出现一次
\usepackage{multirow,colortbl}
\usepackage{subcaption}
\usepackage{amsmath,amssymb}
\usepackage{anyfontsize}
\usepackage{fontawesome5}
\usepackage{algorithm,algorithmic,newfloat}
\usepackage{tcolorbox}
\usepackage{enumitem}
\usepackage{url}
\usepackage{listings}
\usepackage{circledsteps}
\usepackage{pifont}
\usepackage{hyperref}          % **一定最后加载**

\definecolor{darkgreen}{RGB}{0, 100, 0}         % Dark Green
\definecolor{lightgreencustom}{RGB}{220, 255, 220} % Very Light Green
\definecolor{deepskyblue}{RGB}{0,0,0}
\definecolor{crimsonred}{RGB}{220,20,60} % Crimson red

\ifCLASSINFOpdf
\else
\fi
\begin{document}
\title{SV-TrustEval-C: Evaluating Structure and Semantic Reasoning in\\Large Language Models for Source Code Vulnerability Analysis}

% \author{\IEEEauthorblockN{Anonymous Authors}}

% \author{\IEEEauthorblockN{Michael Shell}
% \IEEEauthorblockA{School of Electrical and\\Computer Engineering\\
% Georgia Institute of Technology\\
% Atlanta, Georgia 30332--0250\\
% Email: http://www.michaelshell.org/contact.html}
% \and
% \IEEEauthorblockN{Homer Simpson}
% \IEEEauthorblockA{Twentieth Century Fox\\
% Email: homer@thesimpsons.com}
% \and
% \IEEEauthorblockN{James Kirk\\ and Montgomery Scott}
% \IEEEauthorblockA{Starfleet Academy\\
% San Francisco, California 96678-2391\\
% Telephone: (800) 555--1212\\


% compsoc conferences compared to traditional conferences), use this
\author{
\IEEEauthorblockN{
Yansong Li\IEEEauthorrefmark{1},
Paula Branco\IEEEauthorrefmark{1},
Alexander M. Hoole\IEEEauthorrefmark{2},
Manish Marwah\IEEEauthorrefmark{2},\\
Hari Manassery Koduvely\IEEEauthorrefmark{2},
Guy-Vincent Jourdan\IEEEauthorrefmark{1},
Stephan Jou\IEEEauthorrefmark{2}
}
\IEEEauthorblockA{\IEEEauthorrefmark{1}University of Ottawa, \IEEEauthorrefmark{2}OpenText}
\IEEEauthorblockA{Email: \{yli627, pbranco, gjourdan\}@uottawa.ca, \{hoole, mmarwah, hkoduvely, sjou\}@opentext.com}}

% \IEEEauthorblockA{\IEEEauthorrefmark{2}OpenText}}

% \\
% Email: \{yli627, pbranco, gjourdan\}@uottawa.ca}
% \IEEEauthorblockA{\IEEEauthorrefmark{2}OpenText\\
% Email: \{hoole, mmarwah, hkoduvely, sjou\}@opentext.com}
% }


\maketitle


\begin{abstract}
As Large Language Models (LLMs) evolve in understanding and generating code, accurately evaluating their reliability in analyzing source code vulnerabilities becomes increasingly vital. While studies have examined LLM capabilities in tasks like vulnerability detection and repair, they often overlook the importance of both structure and semantic reasoning crucial for trustworthy vulnerability analysis. To address this gap, we introduce \textsc{SV-TrustEval-C}, a benchmark designed to evaluate LLMs' abilities for vulnerability analysis of code written in the C programming language through two key dimensions: structure reasoning—assessing how models identify relationships between code elements under varying data and control flow complexities; and semantic reasoning—examining their logical consistency in scenarios where code is structurally and semantically perturbed. Our results show that current LLMs are far from satisfactory in understanding complex code relationships and that their vulnerability analyses rely more on pattern matching than on robust logical reasoning. These findings underscore the effectiveness of the \textsc{SV-TrustEval-C} benchmark and highlight critical areas for enhancing the reasoning capabilities and trustworthiness of LLMs in real-world vulnerability analysis tasks. Our initial benchmark dataset is available at \textcolor{blue}{\url{https://huggingface.co/datasets/LLMs4CodeSecurity/SV-TrustEval-C-1.0}}

\end{abstract}

% \fi
\IEEEpeerreviewmaketitle

\section{Introduction}
With the continued development of Large Language Models (LLMs)~\cite{brown2020language, achiam2023gpt, touvron2023llama}, specialized versions tailored for the code domain \cite{roziere2023code, deepseek-coder, bai2023qwen}, have demonstrated promising capabilities in code understanding and generation. Notable applications such as Amazon CodeWhisperer and GitHub Copilot have significantly boosted developer productivity and been widely adopted by millions of developers~\cite{copilotX}. Building on these advances, numerous studies have broadened the applications of LLMs to include cybersecurity tasks, particularly in code vulnerability analysis~\cite{zhang2023pre,ding2024vulnerability,fu2024aibughunter}. Unlike traditional tools that depend on post-hoc verification through static~\cite{tal2004syntax} or dynamic analysis~\cite{ko1994automated}, LLMs offer real-time security feedback during the coding process~\cite{barke2023grounded} and can generate secure code while following detailed instructions~\cite{he2023large}, significantly reducing the risk of software vulnerabilities.

Despite their potential, studies have shown that LLMs do not guarantee code security and are prone to generating syntactically correct but semantically insecure code~\cite{pearce2022asleep,khare2023understanding,Snyk2024}. This raises concerns about their utility in code vulnerability analysis and underscores the urgent need to assess their capabilities in the security domain~\cite{asare2023security,fu2023security}. However, existing benchmarks for code vulnerability, such as CVEFixes~\cite{nikitopoulos2021crossvul}, BigVul~\cite{fan2020ac}, and DiverseVul~\cite{chen2023diversevul}, primarily focus on straightforward tasks like identifying or repairing publicly known vulnerable code snippets within predefined scenarios. These benchmarks fall short as they fail to discern whether success in resolving vulnerabilities stems from leveraging pre-trained parameterized knowledge~\cite{bubeck2023sparks,zhong2023agieval}, or from a logical reasoning between code and vulnerabilities. This distinction is vital for verifying the reliability of LLMs, as reliance exclusively on pre-trained patterns often leads to poor generalization~\cite{huang-chang-2023-towards} and inconsistent interoperability~\cite{ji2023survey}, especially when addressing semantic and structural variations in code within real-world scenarios. To overcome these limitations, we formally define the core capabilities required for vulnerability analysis and introduce \textsc{SV-TrustEval-C}, a new question-answering benchmark designed for the trustworthy evaluation of LLMs in source code vulnerability analysis. This benchmark involves: 

\noindent
\textbf{Structure Reasoning:} Unlike unstructured natural language, code inherently possesses rigorous structural information. For natural language-driven LLMs, the ability to comprehend this code structure is essential for effective code understanding and reasoning~\cite{chen2024reasoning, liu2024codemind}. In vulnerability analysis, understanding the relationships between code elements is critical for assessing the scope and impact of vulnerabilities and for implementing accurate repair strategies. This benchmark evaluates the ability of LLMs to identify and predict the effects of modifications within code elements on each other.

\noindent
\textbf{Semantic Reasoning:} Leveraging domain knowledge to adapt to changes in code is crucial for maintaining robustness in dynamic environments. This benchmark component evaluates LLMs through scenarios that simulate real-world conditions, including: \textbf{Counterfactual scenarios}, where altered code semantics challenge LLMs to apply logical reasoning beyond learned patterns; \textbf{Goal-driven scenarios}, where LLMs are tasked with completing code while ensuring functionality and security, testing their ability to handle complex modifications without introducing vulnerabilities; and \textbf{Predictive scenarios}, where LLMs must accurately identify and differentiate between various types of vulnerabilities, including scenarios where vulnerabilities will not be triggered at runtime, thereby assessing their application of security concepts.

To develop \textsc{SV-TrustEval-C}, we created a Structure-Oriented Variants Generator capable of extracting structural information, perturbing code semantics based on the provided base code and classifications (\textit{Safe}, \textit{Unsafe} code)\footnote{\textit{Safe} and \textit{Unsafe}: defined by whether the code triggers a vulnerability.}, and increasing code complexity in alignment with data flow and control flow graphs~\cite{yamaguchi2014modeling}. We conducted extensive experiments using the \textsc{SV-TrustEval-C} benchmark with zero-shot and in-context learning inference to evaluate eleven popular LLMs across various parameter scales. The results reveal that LLMs struggle to recognize relationships between code elements and predominantly rely on pattern matching rather than logical reasoning in vulnerability analysis scenarios. This highlights the need for specialized improvements in LLMs to enhance their practical utility in real-world code vulnerability analysis. Our contributions are the following:
\begin{enumerate}
    \item \textbf{Reasoning-based Benchmark for Vulnerability Analysis:} We present \textsc{SV-TrustEval-C}, the first benchmark designed to assess LLMs' ability to analyze source code vulnerabilities through logical reasoning and structural understanding, moving beyond mere pattern recognition.
    \item \textbf{Structure-Oriented Variants Generator:} We created a generator that systematically extracts structural information, alters code semantics, and increases complexity based on data and control flow graphs using \textit{Safe} and \textit{Unsafe} code pairs.
    \item \textbf{Identifying Gaps in LLM Capabilities:} Evaluating eleven LLMs revealed their reliance on pattern matching over logical reasoning in vulnerability analysis, highlighting the need for enhancing their reasoning capabilities in security applications.
\end{enumerate}

\begin{table*}[ht]
\centering
\caption{Comparison of recent code vulnerability evaluation benchmarks against our proposed \textsc{SV-TrustEval-C}. “CWE Scope” indicates the number of CWEs in a benchmark; “Num. of Func.” shows the total functions evaluated. Icons indicate dataset origin—\faChild for manually labeled real‑world data, \faCog for automatically labeled real‑world data, and \faEdit for synthetic data. Support levels are shown as \checkmark\ (full), \textcircled{$\checkmark$} (partial), and \ding{56} (none). Key features of our benchmark—CWE coverage, function count, and reasoning capability—are \textcolor{BrickRed}{highlighted} relative to \textsc{SecLLMHolmes}. Evaluation tasks: \textbf{Identification} (Eq.~\ref{eq:bin_dection} and Eq.~\ref{eq:detection}) measures vulnerability detection and localization; \textbf{Repair} (Eq.~\ref{eq:repair}) assesses conversion of vulnerable code into secure versions; \textbf{Generation} (Eq.~\ref{eq:generation}) evaluates producing safe code without introducing new vulnerabilities; \textbf{QA} (Eq.~\ref{eq:QA}) tests domain knowledge via targeted question‑answering; and \textbf{Reasoning} (Eq.~\ref{eq:Reasoning}) evaluates the underlying reasoning process.}

\resizebox{\textwidth}{!}{
\begin{tabular}[width=\textwidth]{@{}lcccccccc@{}}
\toprule
\textbf{Benchmark} & \textbf{CWE Scope} & \textbf{Num. of Func.} & \textbf{Source} & \textbf{Identification} & \textbf{Repair} & \textbf{Generation} & \textbf{QA} & \textbf{Reasoning} \\
\midrule
SVEN~\cite{he2023large} &  9 &  1,606 &  \faChild & \ding{56} & \ding{56} & \checkmark &\ding{56} &\ding{56}\\
Devign~\cite{zhou2019devign} & N/A& 26,037& \faChild & \checkmark & \checkmark & \ding{56} &\ding{56} &\ding{56}\\
VulPatchPairs~\cite{risse2023limits}  & N/A & 26,000& \faChild & \checkmark & \checkmark & \ding{56} &\ding{56} &\ding{56}\\
BigVul~\cite{fan2020ac} & 91 & 264,919 & \faCog  & \checkmark & \checkmark & \ding{56} &\ding{56} &\ding{56} \\
CrossVul~\cite{nikitopoulos2021crossvul} &  168& 134,126 &\faCog  & \checkmark & \checkmark & \ding{56} &\ding{56} &\ding{56} \\
CVEFixes~\cite{nikitopoulos2021crossvul}  & 180& 168,089 & \faCog & \checkmark & \checkmark & \ding{56} &\ding{56} &\ding{56} \\
DiverseVul~\cite{chen2023diversevul} & 150&330,492& \faCog & \checkmark & \checkmark & \ding{56} &\ding{56} &\ding{56} \\
Juliet~\cite{meade2012juliet} & 118 & 61,387 & \faEdit & \checkmark & \checkmark & \ding{56} &\ding{56} &\ding{56}\\
ROBUSTAPI~\cite{zhong2024can} &N/A &1,208&   \faEdit & \ding{56} &\ding{56} & \checkmark & \checkmark  &\ding{56}\\
CyberMetric~\cite{tihanyi2024cybermetric} &N/A&N/A & \faEdit & \ding{56} & \ding{56} & \ding{56} & \checkmark  &\ding{56}\\
CyberBench~\cite{alam2024ctibench}  &N/A&N/A & \faEdit & \checkmark & \ding{56} & \ding{56} & \checkmark  &\ding{56}\\
CYBERSECEVAL2~\cite{bhatt2024cyberseceval}&N/A&N/A  & \faEdit & \ding{56} & \ding{56} & \checkmark & \ding{56}  &\ding{56}\\
\midrule
SecLLMHolmes~\cite{ullah2024llms}  & 8 & 228 & \faChild & \checkmark & \ding{56} & \ding{56} & \ding{56}  & \textcircled{$\checkmark$}\\
\textsc{SV-TrustEval-C} (\textit{Ours}) & \textcolor{BrickRed}{\textbf{82}} &  \textcolor{BrickRed}{\textbf{3,337}} & \faEdit & \checkmark & \ding{56} & \ding{56} & \ding{56}  & \textcolor{BrickRed}{\checkmark}\\
\bottomrule
\end{tabular}}
\label{tab:benchmark_info}
\end{table*}

\section{Background and Related Work} 
% \subsection{Code Vulnerability Analysis} Source code vulnerabilities are weaknesses that attackers can exploit to compromise a system's integrity, availability, or confidentiality~\cite{krsul1998software, kim2012comparative}. 
Historically, identifying and mitigating source code vulnerabilities involved manual code reviews, which were both time-consuming and error-prone~\cite{liu2012software}. Automated tools and methodologies, such as static analysis (e.g., FindBugs~\cite{ayewah2008using}, PMD~\cite{copeland2005pmd}, and Checkstyle~\cite{burn2003checkstyle}) and dynamic analysis techniques like fuzz testing~\cite{klees2018evaluating}, have evolved to increase detection accuracy and efficiency~\cite{harer2018automated, ball1999concept}. Advances in machine learning and deep learning have significantly enhanced vulnerability detection capabilities~\cite{harer2018automated,bilgin2020vulnerability,li2018vuldeepecker,li2021sysevr,chakraborty2021deep,wartschinski2022vudenc}. More recently, LLMs~\cite{brown2020language} have been explored for their potential to understand complex code patterns, predict vulnerabilities, and even automate vulnerability repairs~\cite{feng2020codebert, roziere2023code, guo2024deepseek, nong2024vgx, de2024enhanced, ferrag2024generative}. Despite these advances, questions remain regarding the reliability of LLMs in detecting and repairing specific vulnerabilities under diverse conditions, highlighting the need for rigorous benchmarking to establish their efficacy and trustworthiness in software security applications.

\subsection{Code Vulnerability Benchmarks}
Benchmarks in the code vulnerability domain are designed to evaluate various stages and aspects of how LLMs and other automated tools handle vulnerabilities in code. Broadly, these benchmarks can be categorized according to five key tasks: \textbf{Identification}, \textbf{Repair}, \textbf{Safe-Generation}, \textbf{Question-Answering (QA)} and \textbf{Reasoning}. Each category addresses a distinct component of understanding and managing code vulnerabilities, providing specific datasets and metrics to assess model performance. Table~\ref{tab:benchmark_info} highlights these benchmarks, detailing their focus and labels which of the five tasks they each satisfy. To clearly define the tasks and distinguish our work from existing benchmarks, we formally characterize each of the five vulnerability domain tasks using mathematical formulations in the following sections.

\subsubsection{Identification Tasks}
% These tasks involve detecting and classifying vulnerable code snippets and pinpointing specific vulnerable code segments within these snippets. 

These tasks involve detecting and classifying vulnerable code snippets while evaluating whether each snippet contains vulnerabilities and pinpointing the specific vulnerable segments when necessary. Benchmarks such as \textit{Devign}~\cite{zhou2019devign} and \textit{VulPatchPairs}~\cite{risse2023limits} aim to evaluate a model’s capability to accurately identify whether a piece of code is vulnerable. Additionally, benchmarks like \textit{BigVul}~\cite{fan2020ac}, \textit{CrossVul}~\cite{nikitopoulos2021crossvul}, \textit{CVEFixes}~\cite{nikitopoulos2021crossvul}, and \textit{DiverseVul}~\cite{chen2023diversevul} not only classify code snippets as vulnerable or safe but also provide insights into different types of vulnerabilities gleaned from real-world open-source repositories. Formally, given a code snippet $X$, the objective is usually to predict a binary label $y$ indicating vulnerability:
\begin{align}
    P(y \mid X), \quad \text{where } y \in \{0,1\}, \label{eq:bin_dection}
\end{align}
Some benchmarks extend this task to identify specific lines or elements of code that are vulnerable, which can be formulated as:
\begin{align}
    P(Y \mid X) = \prod_{t=1}^{|X|} P(y_t \mid X, y_{<t}), \label{eq:detection}
\end{align}
where $y_{<t}$ denotes the vulnerability predictions of all preceding lines, and $Y=\{y_1,\dots,y_{|X|}\}$ is the set of predicted labels for each line in $X$. These tasks challenge models to combine global contextual understanding with fine-grained local analysis, reflecting the complex nature of real-world vulnerability detection.

\subsubsection{Repair Tasks}
% Repair tasks focus on generating safe, corrected versions of vulnerable code snippets. 
Repair tasks generate secure versions of vulnerable code snippets by accurately identifying and fixing the underlying issues. Benchmarks that combine identification and repair tasks provide pairs of vulnerable code snippets ($X_v$) and their corresponding fixed versions ($X_r$). Examples include \textit{BigVul}, \textit{CrossVul}, \textit{CVEFixes}, and \textit{DiverseVul}, which gather data from commit records on open-source platforms like GitHub. This data includes the original vulnerabilities and their subsequent fixes, serving as ground truth for evaluating a model's repair capabilities. The formal objective for vulnerability repair tasks is:
\begin{align}
    P(X_r \mid X_v) = \prod_{t=1}^{|X_r|} P(x_t \mid x_{<t}, X_v),\label{eq:repair}
\end{align}

where $X_v$ is the vulnerable code and $X_r$ is the repaired code. This formulation assesses a model’s proficiency in transforming vulnerable code into a secure version, reflecting real-world scenarios where developers or automated tools must identify and rectify security flaws efficiently.

\subsubsection{Safe-Generation Tasks}
These tasks focus on generating secure code snippets while evaluating a model's ability to either produce only safe code when prompted or deny requests that could lead to insecure code, ensuring that no new vulnerabilities are introduced. Benchmarks such as \textit{SVEN}~\cite{he2023large} emphasize controlled code generation, where the LLM generates code only if the input prompt ($c$) is considered "safe," and outputs a denial response if it’s not. This behavior is expressed as:
\begin{align}
    P(X \mid c) = 
    \begin{cases}
    \prod_{t=1}^{|X|} P(x_t \mid x_{<t}, c), & \text{if } c \text{ is safe}; \\
    \delta(X = D), & \text{if } c \text{ is not safe},
    \end{cases} \label{eq:generation}
\end{align}

where the Dirac delta function $\delta(X = D)$ represents a probability distribution concentrated entirely on a denial response $D$. Benchmarks like \textit{RobustAPI}~\cite{zhong2024can} assess how frequently the code generated by an LLM contains vulnerabilities by measuring:
$
P(V(X) = 1 \mid c) = \mathbb{E}_{X \sim P(X \mid c)}[V(X)]$,
where $V(X)$ is an indicator function that equals 1 if $X$ is vulnerable, and 0 otherwise. Other benchmarks, such as \textit{CYBERSECEVAL 2}~\cite{bhatt2024cyberseceval}, investigate controlled code generation in the presence of prompt injection attacks~\cite{liu2023prompt,piet2024jatmo}, measuring the resilience of LLMs to malicious instructions. Collectively, these safe-generation tasks and benchmarks not only evaluate the correctness and functionality of the generated code but also its security and the LLM’s robustness against introducing vulnerabilities.

\subsubsection{Question-Answering Tasks}

Question-Answering (QA) tasks tests the LLM model’s knowledge and understanding of vulnerability concepts by requiring it to accurately and thoroughly answer security related questions with contextually appropriate explanations. Benchmarks like \textit{CyberBench}~\cite{alam2024ctibench} and \textit{CyberMetric}~\cite{tihanyi2024cybermetric} present various types of vulnerability-related questions and require models to provide correct and contextually appropriate answers:
\begin{align}
    P(A \mid Q), \label{eq:QA}
\end{align}
% \[
% $P(A \mid Q)$,
% \]
where $Q$ is a question and $A$ is the model’s answer. These tasks evaluate the depth and applicability of a model’s understanding of vulnerabilities, including its capacity to explain vulnerability concepts, implications, and potential mitigations or fixes.

\subsubsection{Vulnerability Reasoning Tasks} 
Vulnerability Reasoning tasks evaluate a model's ability to logically explain vulnerabilities, differentiating true analytical reasoning from mere pattern matching. While earlier tasks primarily assess an LLM's effectiveness in detecting vulnerabilities, they typically do not clarify whether the model relies on superficial pattern recognition or engages in genuine logical reasoning. This distinction is crucial, as vulnerability analysis demands deep code understanding to accurately interpret complex structures, dependencies, and contextual nuances beyond pattern matching capabilities. Logical reasoning empowers LLMs to identify novel and intricate security flaws, adapt effectively to emerging threats, and deliver accurate, context-sensitive evaluations essential for securing complex software systems. Addressing this critical gap, \textit{SecLLMHolmes}~\cite{ullah2024llms} introduces an approach specifically designed to assess reasoning capabilities in code vulnerability analysis. Vulnerability Reasoning involves formulating both the reasoning process and the identification of vulnerabilities as:
\begin{align}
    P(O \mid X) = \prod_{t=1}^{|O|} P(o_t \mid o_{<t}, X), \label{eq:Reasoning}
\end{align}
where $O = R \, \| \, V$ represents the concatenation of reasoning steps $R$ and vulnerability identification $V$. This task evaluates an LLM's capacity to understand and explain the logic behind a vulnerability. However, \textit{SecLLMHolmes} provides only a limited set of code scenarios (48 hand-crafted code samples) within MITRE’s Top 25 Common Weakness Enumeration (CWE)\footnote{https://cwe.mitre.org/top25/}, %\footnote{Common Weakness Enumeration (CWE) refers to a categorization of common source code vulnerabilities.}
and its evaluation relies primarily on sentence similarity, $\text{Similarity}(O_{\text{LLM}}, O_{\text{Human}})$, between LLM-generated reasoning and human-crafted reasoning. This method doesn't conclusively show if LLMs rely on pattern matching or genuine reasoning due to limited tests and low interpretability.

In this work, we propose a systematic approach to evaluating LLMs' reasoning abilities in vulnerability analysis scenarios. Our benchmark, built upon the synthetically generated Juliet Test Suite covering 82 distinct CWEs, ensures 100\% label accuracy while introducing significantly scalable task complexity to comprehensively evaluate LLMs' vulnerability reasoning capabilities. It is designed to overcome the limitations of existing benchmarks by offering extensive test scenarios and robust evaluation metrics that can better discern genuine logical reasoning from pattern recognition.

\subsection{Data Sources and Quality}
Existing benchmarks utilize data and labels from three primary sources. First, open-source repositories are used in an automated manner, where datasets like \textit{BigVul} and \textit{CrossVul} are created by scraping code and commit information from platforms such as GitHub, with vulnerability labels automatically assigned based on commit metadata and code diffs. While this method can produce extensive datasets, it often suffers from imprecise labeling and inconsistent data quality~\cite{ding2024vulnerability}. Second, some benchmarks, including \textit{Devign} and \textit{VulPatchPairs}, combine automated data collection with manual verification by human experts, ensuring greater label accuracy and more reliable ground truth information, although resulting in smaller datasets. Third, synthetic sources such as the \textit{Juliet Test Suite}\cite{meade2012juliet} and \textit{Romeo}\cite{brust2023romeo} systematically generate code with high label accuracy, though risks of oversimplification remain. Our benchmark dataset is built on synthetic data but extends its complexity through our Structure-Oriented Variants Generator, ensuring both label accuracy and scalable complexity that approximates real-world scenarios.

\subsection{Code Reasoning} Code reasoning involves analyzing and predicting a program's behavior without direct execution, aiding tasks such as vulnerability detection, code generation, and program comprehension. Traditional approaches leverage code execution information—like inputs, outputs, and intermediate runtime states—to enhance model performance in these areas~\cite{henkel2018code, ni2023lever,chen2023teaching, liu2023code}. For example, \textit{Lever}~\cite{ni2023lever} integrated a verifier that utilizes execution results to improve code generation performance, while Chen et al.\cite{chen2023teaching} employed output-based feedback for guiding LLMs in self-debugging generated code. Meanwhile, some studies have incorporated dynamic features such as runtime program states to train language models capturing deeper code semantics~\cite{liu2023code, ding2024traced}. Recent benchmarks have started evaluating the code reasoning abilities of LLMs. CodeMind~\cite{liu2024codemind} tests how well models reason about input-output relations in code, whereas CRUXEval~\cite{chen2024reasoning} goes further to measure how models infer runtime behavior, including code coverage and execution paths. These benchmarks demonstrate progress in assessing LLMs' code understanding but primarily focus on general code correctness and runtime behavior inference.

Our work differs by explicitly emphasizing code vulnerability reasoning. Unlike general code reasoning tasks that concentrate on functionality and runtime behavior, vulnerability reasoning requires understanding complex code semantics and contextual security considerations. Existing benchmarks like CRUXEval and CodeMind do not focus on this specialized aspect. We introduce a benchmark tailored to evaluating how well LLMs handle vulnerability reasoning tasks, aiming to provide more targeted insights into model capabilities in analyzing code for security flaws and advancing our understanding of code reasoning in security contexts.

\begin{figure*}[th]
    \centering
\includegraphics[width=\textwidth]{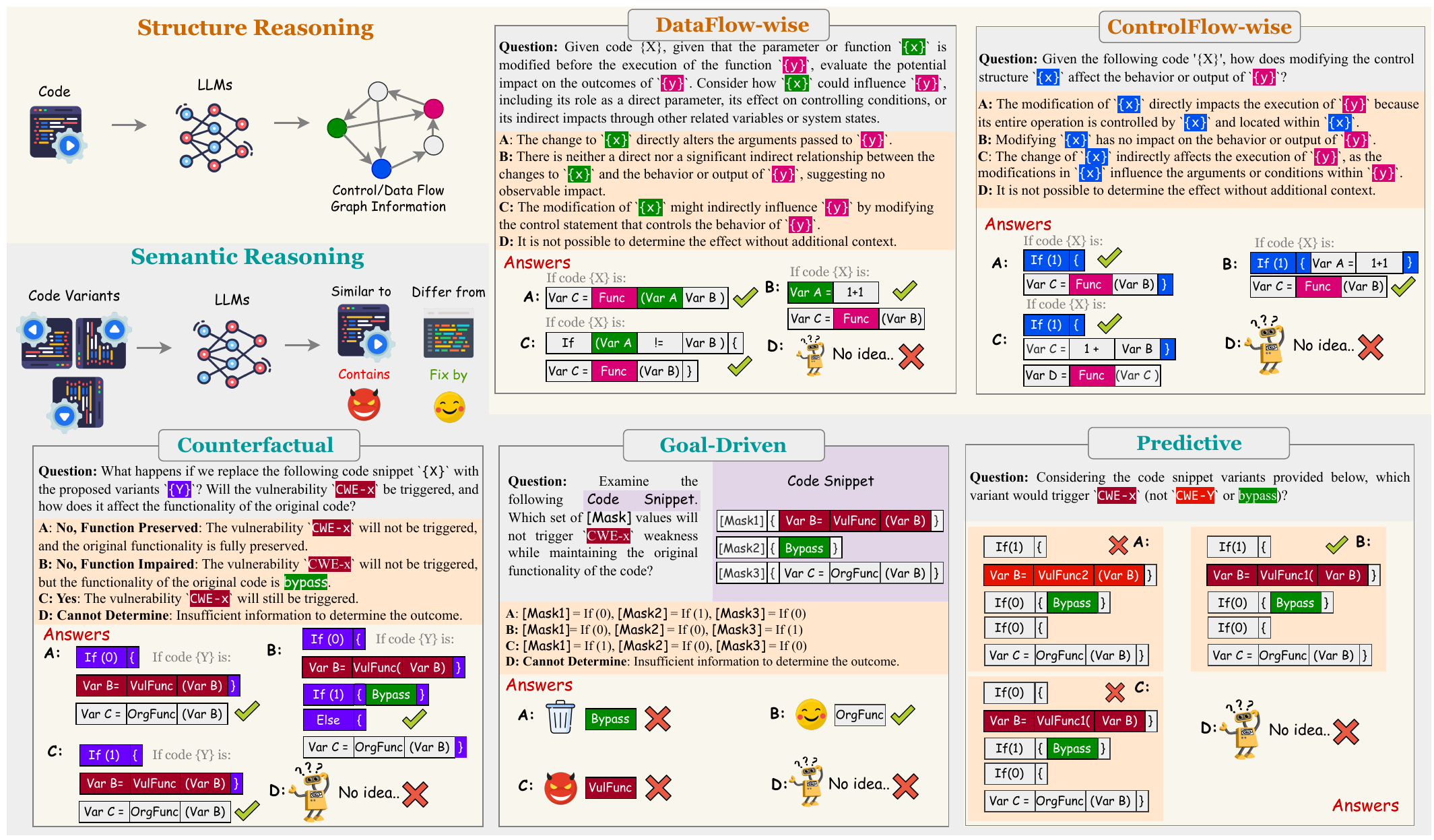}
    \caption{Overview of the \textsc{SV-TrustEval-C} Benchmark: This figure highlights question-answering templates for \textbf{Structure Reasoning}, divided into control-flow and data-flow segments, alongside \textbf{Semantic Reasoning}, encompassing Counterfactual, Goal-driven, and Predictive tasks. Each section illustrates various question types and potential answer scenarios, designed to assess LLMs’ abilities to handle the diverse complexities of code semantics and structure.}
    \label{fig:QA_overview}
\end{figure*}

% \section{Related Work}

\section{SV-TrustEval-C Benchmark}
This section first outlines the capabilities necessary for reliable vulnerability analysis. Subsequently, we describe the key features of the \textsc{SV-TrustEval-C} Benchmark, which leverages our Structure-Oriented Variants Generator.

\subsection{Core Capabilities}
To thoroughly evaluate the reliability of LLMs in vulnerability analysis, we design two main categories of tasks: \textbf{Structure Reasoning} and \textbf{Semantic Reasoning}. These categories correspond to fundamental aspects of code vulnerability tasks, as illustrated in Figure~\ref{fig:QA_overview}.

\subsubsection{Structure Reasoning}
Structure Reasoning evaluates how accurately LLMs understand the relationships and interactions between code elements, focusing on both data flow and control flow, a critical capability for identifying and mitigating potential security threats. This category aligns with the task:
\begin{align}
    P(D \mid X) = \prod_{i=1}^{M} \prod_{j=1}^{M} P(D_{ij} \mid X),\label{eq:SR_Eq}
\end{align}
% \[
% P(D \mid X) = \prod_{i=1}^{M} \prod_{j=1}^{M} P(D_{ij} \mid X),
% \]
where $D_{ij}$ denotes the relationship between elements $i$ and $j$ (with $D_{ij}=1$ indicating a connection, and $0$ otherwise), $M$ is the total number of elements, and $X$ represents the code snippet. Specifically:
\begin{itemize}
    \item \textbf{DataFlow-wise Reasoning}: Evaluates the model's understanding of how data moves through the code, essential for identifying vulnerabilities related to data handling.
    \item \textbf{ControlFlow-wise Reasoning}: Assesses the model's proficiency in analyzing the control flow of the program, vital for understanding how different parts of the code execute and interact.
\end{itemize}
These tasks examine the model's ability to discern how code elements correlate and how vulnerabilities can propagate through these interactions.

\subsubsection{Semantic Reasoning}
% Semantic Reasoning evaluates an LLM's adaptability and understanding of changes in code semantics under various scenarios. This category encompasses three sub-tasks: \textbf{Counterfactual}, \textbf{Goal-driven}, and \textbf{Predictive}. Counterfactual reasoning tests how well an LLM can predict vulnerabilities when code has been altered. 

% Semantic Reasoning tests how well an LLM handles code transformations, ensuring security and functionality. It spans three sub-tasks: Counterfactual (predicting how changes affect vulnerabilities), Goal-driven (safely modifying code to meet specific aims), and Predictive (classifying code variants by their security impact).

Semantic Reasoning evaluates an LLM's adaptability and understanding of changes in code semantics under various scenarios and transformations, ensuring security and functionality, and encompasses three sub-tasks: \textbf{Counterfactual} (predicting vulnerabilities when code is altered), \textbf{Goal-driven} (safely modifying code to meet specific aims), and \textbf{Predictive} (classifying code variants by their security impact). Formally, for the Counterfactual scenario, we consider:
\begin{equation}\label{eq:Counterfactual_eq}
\begin{split}
    P\bigl(V(f(X_s, t)) = 1 \mid f(X_s, t), X_s\bigr),\\
    t \in \{\text{safe},\,\text{unsafe},\,\text{impaired}\}
\end{split}
\end{equation}
where $V(X)=1$ if $X$ is vulnerable (otherwise $0$), and $f(X_s, t)$ transforms $X_s$ based on behavior $t$. We introduce three transformation types:
\begin{itemize}
    \item \textbf{Safe Code Transformations}: Modifications that preserve functionality without adding vulnerabilities.
    \item \textbf{Unsafe Code Transformations}: Modifications that introduce new vulnerabilities into the code.
    \item \textbf{Impaired Code Transformations}: Alterations that keep the code vulnerability-free but diminish its original functionality.    
\end{itemize}
These tasks assess the model's consistency and adaptability in analyzing vulnerabilities under semantic alterations, reinforcing the importance of understanding how code transformations affect security properties. 

Goal-driven tasks evaluate an LLM's ability to modify code to achieve a specific outcome without introducing vulnerabilities:
\begin{align}
    P(V(X^{}) = 0, X \mid c), \label{eq:Goal-driven_eq}
\end{align}
% \[
% $P(V(X^{}) = 0, X \mid c)$,
% \]
where $X$ is the modified code snippet to meet the goal, $c$ is the given context or constraint, and $V(X) = 0$ means the modified code remains vulnerability-free. In these tasks:
\begin{itemize}
    \item The model is prompted to insert or alter code to achieve specified goals (e.g., adding features or fixing bugs) while ensuring no new vulnerabilities are introduced.
    \item Code templates emphasize control statements that can either bypass, trigger, or prevent vulnerabilities.
\end{itemize}
By ensuring the resulting code remains secure and functional, these scenarios evaluate the model's proficiency in context-aware vulnerability reasoning and code refinement. The joint probability formulation, $P(V(X^{}) = 0, X \mid c)$, assesses both the safety of the modified code and its successful production under the given context $c$. 

Predictive scenarios challenge the LLM to classify code variants based on whether they introduce, remove, or do not affect vulnerabilities and code functionality:
\begin{align}
    P(k \mid f(X)), \label{eq:Predictive_eq}
\end{align}
%\[
% $P(k \mid f(X))$,
% \]
where $k$ represents the code state or type of vulnerability, including non-vulnerable and potentially impaired states, and $f(X)$ is the code variants. These tasks:
\begin{itemize}
    \item Provide code variants that may introduce new vulnerabilities, remove existing vulnerabilities, remain vulnerability-free but become functionally impaired, or have no effect on vulnerability.
    \item Require the LLM to accurately classify these code variants, distinguishing among various vulnerability states and the possibility of impaired functionality.
\end{itemize}
These Predictive tasks assess the model's capability to distinguish among different vulnerabilities and its understanding of how code modifications influence security.

\begin{figure}[t]
    \centering
    \includegraphics[width=0.35\textwidth]{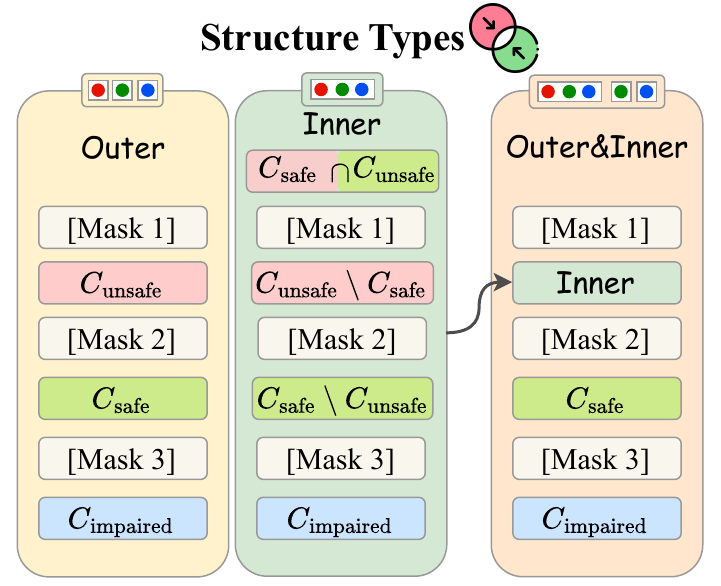}
    \caption{Overview of the Outer, Inner, and Outer\&Inner structures. The \textit{impaired} block prints or returns \textit{None} and does not mimic the behavior of either $C_{\text{safe}}$ or $C_{\text{unsafe}}$. Here, $C_{\text{safe}}\cap C_{\text{unsafe}}$ represents the common parts of both versions, while $C_{\text{safe}}\setminus C_{\text{unsafe}}$ denotes the unique parts of $C_{\text{safe}}$.}
    \label{fig:main_method}
\end{figure}
\subsection{Structure-Oriented Variants Generator}
This section describes the purpose of our Structure-Oriented Variants Generator, subsequently referred to as generator, and its role in generating our benchmark dataset. The generator contains a Flow Extractor and a Behaviour Simulator. Given a code snippet from an existing codebase, the generator strategically modifies the behavior of code variants into three categories: \textit{safe} (preventing vulnerabilities), \textit{unsafe} (triggering vulnerabilities), and \textit{impaired} (bypassing the original function), while enhancing their structural complexity. It also scales the structural complexity to suit the benchmark questions, ensuring a diverse set of code variants.

\subsubsection{Flow Extractor}
Given a code snippet $C$, our Flow Extractor constructs the data flow graph $\mathcal{G}_d = (\mathcal{V}_d, \mathcal{E}_d)$ and the control flow graph $\mathcal{G}_c = (\mathcal{V}_c, \mathcal{E}_c)$. Initially, we parse $C$ into an Abstract Syntax Tree (AST) using a parser generator tool\footnote{https://tree-sitter.github.io/tree-sitter/} to extract code syntactic information. We then apply Depth-First Search (DFS) to traverse the AST, identifying code elements $v_i \in \mathcal{V}_d$ and $v_i \in \mathcal{V}_c$ such as variables, literals, and expressions, assuming different contexts for data and control flows respectively. Concurrently, we establish edges $e^{d}_{i,j} \in \mathcal{E}_d$ and $e^{c}_{i,j} \in \mathcal{E}_c$ to represent data dependencies between $v^{d}_i$ and $v^{d}_j$ through assignments and function calls, as well as control flows between elements $v^{c}_i$ and $v^{c}_j$.

\subsubsection{Behaviour Simulator}

Given the graph data extracted by the Flow Extractor and the required behaviors of the code (\textit{safe}, \textit{unsafe}, and \textit{impaired}), our behavior simulator manages the final behavior of the code by modifying the safe code $C_{\text{safe}}$ and the unsafe code $C_{\text{unsafe}}$ within the original code $C$. Initially, the simulator incorporates control flow variants from Juliet~\cite{meade2012juliet}, introducing additional control branches within $\mathcal{G}_c$. This step tests if LLMs can maintain consistency with code variants that have similar syntax but different structures. As illustrated in Figure 2, the simulator then introduces Outer, Inner, and Outer\&Inner structures to manage the code's behavior, infusing $C_{\text{safe}}$, $C_{\text{unsafe}}$ and our $C_{\text{impaired}}$ with masked control statements into various structural configurations. Next, we fill the masked control statements to create code variants with the specified behavior. Finally, the Behavior Simulator generates multiple variants for each base code $C$, categorized by structure and behavior, to test LLM adaptability and accuracy.

\subsection{Question Generation}
After creating code variants from existing codebases, we develop questions for structural and semantic reasoning scenarios. Sample questions are provided in Appendix~\ref{sec:sample}.

\subsubsection{DataFlow-wise Questions}
\begin{figure}[t]
    \centering
\includegraphics[width=0.48\textwidth]{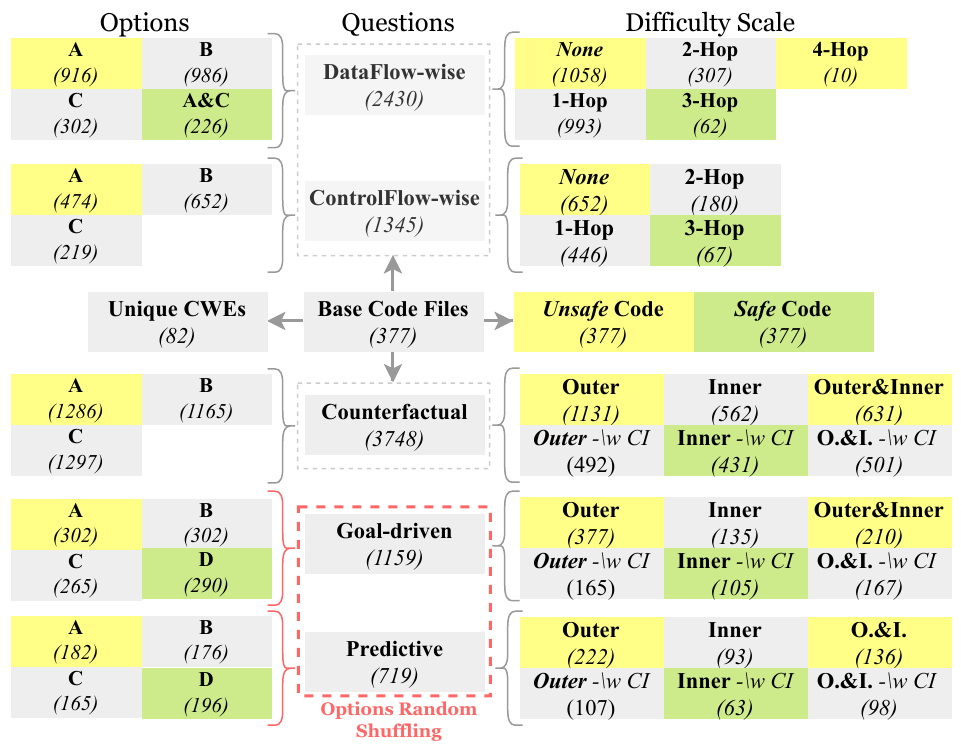}
    \caption{Overview of benchmark question distribution across two core capabilities, five question types, and multiple difficulty levels. The notation \textit{-\textbackslash w CI} indicates code variants with control flow injection. The term \textit{None} indicates that there is no connection within the dataflow or control flow graph in the structure reasoning scenario. The question generation strategy is designed to ensure a uniform distribution across each base file, prioritizing high difficulty levels.}
    \label{fig:data_statistic}
\end{figure}

For DataFlow-wise structure reasoning, we use the Flow Extractor to construct the data flow graph $\mathcal{G}_d$. We then evaluate each source node $v_i^{d} \in \mathcal{V}_d$ together with function-only target nodes $v_j^{d}$ to determine whether modifying $v_i^{d}$ influences $v_j^{d}$’s behavior. Each generated question falls into one of the following categories (see Figure~\ref{fig:data_statistic}): 
\begin{itemize}
    \item \textbf{A:} Ground truth is derived directly from $\mathcal{E}_d$, which may involve direct or multi-hop connections within $\mathcal{G}_d$. 
    \item \textbf{B:} No connection in either $\mathcal{G}_d$ or $\mathcal{G}_c$.
    \item \textbf{C:} Some connections are not explicitly shown in $\mathcal{G}_d$ but are implied through control flow $\mathcal{G}_c$. For instance, if $v_j^{d}$ is within an \texttt{if($v_i^{d}==1$)} block, and $v_i^{d}$ is involved in the condition, it has an indirect impact. Similarly, if $v_i^{d}$ triggers a \texttt{break} that prevents execution from reaching $v_j^{d}$, it also constitutes an indirect impact.
    \item \textbf{A\&C:} In some cases, both A and C apply—namely, when $v_i^d$ is connected to $v_j^d$ in $\mathcal{G}_d$ \textit{and} also influences a control statement governing $v_j^d$.
\end{itemize}
The difficulty of each question increases with the number of hops between $v_i^d$ and $v_j^d$.

\subsubsection{ControlFlow-wise}
For ControlFlow-wise structure reasoning, we use $\mathcal{G}_c$, obtained from the Flow Extractor, to enumerate all control statements $v_i^{c} \in \mathcal{V}_c$ linked to function-only target nodes $v_j$, to assess whether modifying $v_i^{c}$ would impact $v_j^{c}$. Each generated question belongs to one of the following classes/options.
\begin{itemize}
    \item \textbf{A:} The impact of control statement $v_i^{c}$ on function $v_j^{c}$ can be directly obtained from $\mathcal{G}_c$.
    \item \textbf{B:} No perceivable connections within either $\mathcal{G}_d$ or $\mathcal{G}_c$.
    \item \textbf{C:} The control statement $v_i^{c}$ directly or indirectly affects the arguments of $v_j^{c}$ as specified within $\mathcal{G}_d$.
\end{itemize}
The difficulty scale for ControlFlow-wise questions is scaled by the number of hops from $v_i^{c}$ to $v_j^{c}$ within $\mathcal{G}_c$.

\subsubsection{Counterfactual} 
In the Counterfactual semantic reasoning task, for each code snippet $C$, we generate code variants $\mathcal{C}$ using the Structure-Oriented Variants Generator to construct a set of options. Each generated question falls into one of the following classes.
\begin{itemize}
    \item \textbf{A:} The \textit{safe} code variants do not trigger the vulnerability while preserving $C$'s original functionality.
    \item \textbf{B:} The \textit{impaired} code variants avoid triggering the vulnerability but fail to maintain $C$'s functionality.
    \item \textbf{C:} The \textit{unsafe} code variants trigger the vulnerability.
\end{itemize}
The difficulty level for Counterfactual scenarios is scaled based on the number of control flow injections and the specific code structures introduced by the generator.

\subsubsection{Goal-driven}
The goal-driven semantic reasoning questions assess LLMs' ability to fix vulnerabilities without altering code functionality. We generate code variants using the Structure-Oriented Variants Generator with masked control statements as conditions. LLMs are then tasked with selecting the correct path to resolve the vulnerability while preserving the intended functionality. Answer options are generated using the Behavior Simulator and include: (1) \textbf{Bypass}, (2) \textbf{Resolve (Expected Behavior)}, and (3) \textbf{Trigger the Vulnerability}. The difficulty of each question scales with the complexity of the underlying code variant. Since the correct answer is always the same class (i.e., ``Resolve") in this task, we apply \textbf{random shuffling} of both the options and answer placements to prevent position bias.

\begin{table*}[ht]
\centering
\caption{Comparison of model performance based on accuracy. The \colorbox{darkgreen!50}{\textcolor{white}{best performance}} and \colorbox{lightgreencustom}{\textcolor{black}{second-best performance}} are highlighted to denote the top two scores, respectively. The $\star$ symbol identifies models in the ``instruct'' version, which are specifically fine-tuned to enhance instruction-following capabilities. Additionally, we report the false positive rate (FPR) for misclassified ``safe'' instances (FPR$^{\textbf{safe}}$) in the baseline scenario. The \textcolor{crimsonred}{$\downarrow$} symbol indicates a decrease in model performance compared with \texttt{Zero} on the current task after introducing \texttt{ICL}, while the \textcolor{deepskyblue}{$\uparrow$} symbol signifies an improvement in performance.}

\resizebox{\textwidth}{!}{
\begin{tabular}{@{}l|cccccc|cccccccc|cccc}
\toprule
\multirow{3}{*}{\textbf{Models}} & \multicolumn{6}{c|}{\textbf{Structure Reasoning}} & \multicolumn{8}{c|}{\textbf{Semantic Reasoning}} & \multicolumn{4}{c}{\textbf{Base Scenario}}\\

 & \multicolumn{2}{c}{\textbf{DataFlow}} & \multicolumn{2}{c}{\textbf{ControlFlow}} & \multicolumn{2}{c|}{\textbf{Average}} & \multicolumn{2}{c}{\textbf{Counterfactual}} & \multicolumn{2}{c}{\textbf{Goal-driven}} & \multicolumn{2}{c}{\textbf{Predictive}} & \multicolumn{2}{c|}{\textbf{Average}} & \multirow{2}{*}{\textit{\textbf{Unsafe}}} & \multirow{2}{*}{\textit{\textbf{Safe}}} & \multirow{2}{*}{\textbf{Average}} & \multirow{2}{*}{\textbf{FPR$^{\textbf{Safe}}$}}\\
 
 & \texttt{Zero} & \texttt{ICL} & \texttt{Zero} & \texttt{ICL} & \texttt{Zero} & \texttt{ICL} & \texttt{Zero} & \texttt{ICL} & \texttt{Zero} & \texttt{ICL} & \texttt{Zero} & \texttt{ICL} & \texttt{Zero} & \texttt{ICL} \\

\midrule
\multicolumn{19}{c}{\cellcolor{gray!10}$>$\textbf{15B Param. Models}} \\
\midrule
\texttt{GPT-4-turbo} &\cellcolor{lightgreencustom}{60.58} & \cellcolor{lightgreencustom}{59.42}\textcolor{crimsonred}{$\downarrow$} & \cellcolor{lightgreencustom}{65.20} & \cellcolor{lightgreencustom}{68.55}\textcolor{deepskyblue}{$\uparrow$} & \cellcolor{lightgreencustom}{62.89} & \cellcolor{lightgreencustom}{63.98}\textcolor{deepskyblue}{$\uparrow$} & \cellcolor{lightgreencustom}{41.57} & \cellcolor{lightgreencustom}{52.27}\textcolor{deepskyblue}{$\uparrow$} & \cellcolor{lightgreencustom}{31.23} & 1.47\textcolor{crimsonred}{$\downarrow$} & 19.61 & \cellcolor{lightgreencustom}{30.46}\textcolor{deepskyblue}{$\uparrow$} & \cellcolor{lightgreencustom}{30.80} & 28.07\textcolor{crimsonred}{$\downarrow$} & 56.65 & 76.39 & \cellcolor{darkgreen!50}{\textcolor{white}{66.58}} & \cellcolor{darkgreen!50}{\textcolor{white}{23.61}}\\
    
\texttt{GPT-3.5-turbo} &40.41 & 50.49\textcolor{deepskyblue}{$\uparrow$} & 56.95 & 47.36\textcolor{crimsonred}{$\downarrow$} & 48.68 & 48.9\textcolor{deepskyblue}{$\uparrow$} & 32.76 & 34.79\textcolor{deepskyblue}{$\uparrow$} & 4.83 & 16.82\textcolor{deepskyblue}{$\uparrow$} & 19.47 & 23.50\textcolor{deepskyblue}{$\uparrow$} & 19.02 & 25.04\textcolor{deepskyblue}{$\uparrow$} & 56.23 & 31.56 & 43.90 & 68.44\\

\texttt{Llama3.1-405B}$^{\star}$ &\cellcolor{darkgreen!50}{\textcolor{white}{63.17}} & \cellcolor{darkgreen!50}{\textcolor{white}{61.40}}\textcolor{crimsonred}{$\downarrow$} & \cellcolor{darkgreen!50}{\textcolor{white}{73.98}} & \cellcolor{darkgreen!50}{\textcolor{white}{76.88}}\textcolor{deepskyblue}{$\uparrow$} & \cellcolor{darkgreen!50}{\textcolor{white}{68.58}} & \cellcolor{darkgreen!50}{\textcolor{white}{69.14}}\textcolor{deepskyblue}{$\uparrow$} & \cellcolor{darkgreen!50}{\textcolor{white}{45.17}} & \cellcolor{darkgreen!50}{\textcolor{white}{62.57}}\textcolor{deepskyblue}{$\uparrow$} & 2.67 & 29.08\textcolor{deepskyblue}{$\uparrow$} & \cellcolor{lightgreencustom}{28.93} & \cellcolor{darkgreen!50}{\textcolor{white}{43.95}}\textcolor{deepskyblue}{$\uparrow$} & 25.59 & \cellcolor{darkgreen!50}{\textcolor{white}{45.20}}\textcolor{deepskyblue}{$\uparrow$} & 87.00 & 24.14 & 55.57 & 75.86\\

\midrule
\multicolumn{19}{c}{\cellcolor{gray!10}$<$\textbf{15B Param. Models}} \\
\midrule 
\texttt{Llama3.1-8B}$^{\star}$ &45.80 & 54.57\textcolor{deepskyblue}{$\uparrow$} & 54.94 & 52.64\textcolor{crimsonred}{$\downarrow$} & 50.37 & 53.60\textcolor{deepskyblue}{$\uparrow$} & 36.87 & 38.42\textcolor{deepskyblue}{$\uparrow$} & 1.12 & 23.04\textcolor{deepskyblue}{$\uparrow$} & 24.90 & 27.54\textcolor{deepskyblue}{$\uparrow$} & 20.96 & 29.67\textcolor{deepskyblue}{$\uparrow$} & 55.17 & 66.31 & \cellcolor{lightgreencustom}{60.74} & \cellcolor{lightgreencustom}{33.69}\\

\texttt{Llama3-8B}$^{\star}$ &46.63 & 49.79\textcolor{deepskyblue}{$\uparrow$} & 44.68 & 48.48\textcolor{deepskyblue}{$\uparrow$} & 45.66 & 49.14\textcolor{deepskyblue}{$\uparrow$} & 34.82 & 35.33\textcolor{deepskyblue}{$\uparrow$} & 0.95 & 6.04\textcolor{deepskyblue}{$\uparrow$} & 23.92 & 29.49\textcolor{deepskyblue}{$\uparrow$} & 19.9 & 23.62\textcolor{deepskyblue}{$\uparrow$} & 53.85 & 60.74 & 57.29 & 39.26\\

\texttt{CodeLlama-7B}$^{\star}$ 
&38.40 & 37.70\textcolor{crimsonred}{$\downarrow$} & 34.65 & 29.22\textcolor{crimsonred}{$\downarrow$} & 36.52 & 33.46\textcolor{crimsonred}{$\downarrow$} & 34.26 & 34.61\textcolor{deepskyblue}{$\uparrow$} & 0.00 & \cellcolor{lightgreencustom}{33.74}\textcolor{deepskyblue}{$\uparrow$} & 24.48 & 29.07\textcolor{deepskyblue}{$\uparrow$} & 19.58 & 32.47\textcolor{deepskyblue}{$\uparrow$} & 100.00 & 0.00 & 50.00 & 100.00\\

\texttt{CodeLlama-13B}$^{\star}$ &23.74 & 37.57\textcolor{deepskyblue}{$\uparrow$} & 36.43 & 34.35\textcolor{crimsonred}{$\downarrow$} & 30.08 & 35.96\textcolor{deepskyblue}{$\uparrow$} & 34.47 & 34.61\textcolor{deepskyblue}{$\uparrow$} & 24.68 & 23.38\textcolor{crimsonred}{$\downarrow$} & 22.67 & 26.01\textcolor{deepskyblue}{$\uparrow$} & 27.27 & 28.02\textcolor{deepskyblue}{$\uparrow$} & 100.00 & 0.00 & 50.00 & 100.00\\

\texttt{Gemma-7B}$^{\star}$ &36.17 & 33.87\textcolor{crimsonred}{$\downarrow$} & 27.88 & 29.44\textcolor{deepskyblue}{$\uparrow$} & 32.02 & 31.66\textcolor{crimsonred}{$\downarrow$} & 33.11 & 33.62\textcolor{deepskyblue}{$\uparrow$} & 28.90 & \cellcolor{darkgreen!50}{\textcolor{white}{41.59}}\textcolor{deepskyblue}{$\uparrow$} & 24.48 & 26.70\textcolor{deepskyblue}{$\uparrow$} & 28.83 & \cellcolor{lightgreencustom}{33.97}\textcolor{deepskyblue}{$\uparrow$} & 99.73 & 0.53 & 50.13 & 99.47\\

\texttt{CodeGemma-7B}$^{\star}$ &37.98 & 40.00\textcolor{deepskyblue}{$\uparrow$} & 29.29 & 39.85\textcolor{deepskyblue}{$\uparrow$} & 33.64 & 39.92\textcolor{deepskyblue}{$\uparrow$} & 35.01 & 35.17\textcolor{deepskyblue}{$\uparrow$} & 0.09 & 22.78\textcolor{deepskyblue}{$\uparrow$} & 21.56 & 16.97\textcolor{crimsonred}{$\downarrow$} & 18.89 & 24.97\textcolor{deepskyblue}{$\uparrow$} & 98.14 & 0.27 & 49.20 & 99.73\\

\texttt{CodeQwen1.5-7B}$^{\star}$ &35.76 & 38.02\textcolor{deepskyblue}{$\uparrow$} & 34.87 & 35.24\textcolor{deepskyblue}{$\uparrow$} & 35.32 & 36.63\textcolor{deepskyblue}{$\uparrow$} & 33.11 & 32.84\textcolor{crimsonred}{$\downarrow$} & \cellcolor{darkgreen!50}{\textcolor{white}{33.74}} & 22.69\textcolor{crimsonred}{$\downarrow$} & 26.70 & 21.70\textcolor{crimsonred}{$\downarrow$} & \cellcolor{darkgreen!50}{\textcolor{white}{31.18}} & 25.74\textcolor{crimsonred}{$\downarrow$} & 100.00 & 0.00 & 50.00 & 100.00\\

\texttt{Mixtral-7B}$^{\star}$ &29.67 & 51.77\textcolor{deepskyblue}{$\uparrow$} & 37.70 & 52.49\textcolor{deepskyblue}{$\uparrow$} & 33.68 & 52.13\textcolor{deepskyblue}{$\uparrow$} & 31.78 & 32.90\textcolor{deepskyblue}{$\uparrow$} & 0.00 & 9.66\textcolor{deepskyblue}{$\uparrow$} & \cellcolor{darkgreen!50}{\textcolor{white}{29.62}} & \cellcolor{lightgreencustom}{30.46}\textcolor{deepskyblue}{$\uparrow$} & 20.47 & 24.34\textcolor{deepskyblue}{$\uparrow$} & 3.98 & 0.00 & 1.99 & 100.00\\
\toprule
\end{tabular}}
\label{tab:main_res}
\end{table*}

\subsubsection{Predictive}
The Predictive semantic reasoning tasks also use code variants generated by the generator, challenging LLMs to identify which variants trigger specific vulnerabilities. This evaluation tests the models' ability to detect and accurately predict the presence of vulnerabilities, and to differentiate between various types of vulnerabilities. The answer options—(1) \textbf{Bypass}, (2) \textbf{Target CWE (Expected Behavior)}, and (3) \textbf{Different CWE}—follow the same \textbf{random shuffling} strategy as the Goal-driven questions. Difficulty scales with the complexity of each variant.

\subsection{Benchmark Statistics}
The \textsc{SV-TrustEval-C} benchmark, developed using the C programming language components from the Juliet Test Suite~\cite{meade2012juliet}, includes 377 base files, each containing both a \textit{safe} and an \textit{unsafe} function, covering 82 distinct CWEs. Our generator produced a total of 1,297 \textit{unsafe} and 1,286 \textit{safe} compilable code variants. Utilizing these variants, we created 9,401 questions. The detailed questions' statistics are presented in Figure~\ref{fig:data_statistic}.

\section{Experiments}
\subsection{Experimental Setups}
We evaluate eleven of the most popular and representative large language models (LLMs), including \texttt{GPT-4-turbo}~\cite{achiam2023gpt}, \texttt{GPT-3.5-turbo}~\cite{brown2020language}, \texttt{Llama3.1}~\cite{dubey2024llama} in both 405B and 8B versions, \texttt{Llama3}~\cite{llama3modelcard}, \texttt{CodeLlama}\cite{roziere2023code} in 13B and 7B versions, \texttt{Gemma}\cite{team2024gemma}, \texttt{CodeGemma}~\cite{team2024codegemma}, \texttt{CodeQwen}~\cite{bai2023qwen}, and \texttt{Mixtral}~\cite{jiang2024mixtral}. We excluded the latest LLM, \texttt{GPT-o1}~\cite{openai2024o1}, due to budget constraints and prohibitively high time consumption. All models have undergone specialized pre-training in the code domain and are available in the ``instruct" version, which is fine-tuned to follow prompted instructions, making them well-suited for our benchmark design. We conduct all model inferences at a temperature of \textbf{zero} to ensure more deterministic answers and set a maximum output length of 50 tokens, as only the selected option is required. All other inference hyperparameters are set to their default values for each LLM. Furthermore, we establish conversation-independent threads for each question in our experiment to eliminate potential information leakage during the question-answer (QA) process.

\begin{figure}[t]
\centering
\begin{tcolorbox}[
    width=0.48\textwidth,
    boxrule=1pt,
    colback=gray!5,
    colframe=black!75!white,
    title=In-context Learning Prompt,
    fonttitle=\bfseries,
    boxsep=5pt,
    left=5pt,
    right=5pt,
    top=5pt,
    bottom=5pt,
    arc=3pt
]
\textbf{Instruction:} \{Assign a code security expert persona with guidelines to answer the question.\}

\textbf{Context:}
\begin{itemize}[leftmargin=*]
    \item \textbf{Question(s):} \{Insert the demo question based on structural or semantic reasoning.\}
    \item \textbf{Choices:} \{Provide four possible answers based on the code scenario.\}
    \item \textbf{Answer(s):} \{Specify the correct answer, e.g., ``B''.\}
    \item \textbf{Explanation(s):} \{Provide a detailed explanation justifying why the chosen answer is correct.\}
\end{itemize}
\textbf{Input Question:} \{Insert the selected question.\}\\ \textbf{Output:} \{LLM-generated answer.\}
\end{tcolorbox}
\caption{Prompt template for In-context learning pipeline}
\label{llm_prompt_framework}
\end{figure}

% \begin{figure}[t]
% \centering
% \begin{tcolorbox}[
%     width=0.48\textwidth,
%     boxrule=0.5pt,
%     colback=white,
%     colframe=black,
%     title=In-context Learning Prompt,
%     boxsep=5pt,             % Reduces inner separation
%     left=5pt,               % Adjust left padding
%     right=5pt,              % Adjust right padding
%     top=5pt,                % Adjust top padding
%     bottom=5pt              % Adjust bottom padding
% ]

% \textbf{Instruction:} \{Assign a persona specialized in the code security domain and provide specific guidelines for addressing the following question.\}

% \vspace{0.3em} % Reduced vertical space for tightness

% \textbf{Context:}
% \textbf{Question(s):} \{Insert demo question applicable based on structure reasoning or semantic reasoning.\}
% \vspace{0.3em}\\
% \textbf{Choices:} \{Provide four possible answers based on the current code scenario.\}
% \vspace{0.3em}\\
% \textbf{Answer(s):} \{Specify the correct choice, e.g., "B"\}
% \vspace{0.3em}\\
% \textbf{Explanation(s):} \{Provide a detailed, human-level explanation justifying why the chosen answer is correct.\}
% \textbf{Input Question:} \{Selected question from the benchmark\}
% \textbf{Output:} \{LLMs generated answerr\}
% \end{tcolorbox}
% \caption{Prompt template for in-context learning pipeline.}
% \label{llm_prompt_framework}
% \end{figure}
\subsubsection{Inference Mode}
We utilize both zero-shot inference (\texttt{Zero}) and in-context learning (\texttt{ICL}) approaches~\cite{liu2021makes,lu2021fantastically,wu2022self} to comprehensively evaluate LLMs. In the zero-shot setup, models generate responses based solely on the input question without additional examples, allowing us to assess their inherent understanding and reasoning abilities. In contrast, the in-context learning approach provides models with specific prompts that demonstrate question-and-answer patterns along with corresponding explanations for each answer. This contextual information guides their responses, enhancing their ability to learn from context and more accurately follow the question. The prompts used for in-context learning include a few sample question-response pairs following the template shown in Figure~\ref{llm_prompt_framework}. All questions, answer choices, correct answers, and explanations are carefully crafted by human domain experts and subsequently refined by \texttt{GPT-o1} to ensure accuracy and reduce instructional errors.

\subsubsection{Label Masking} To avoid label leakage from the original Juliet test suite, we applied label masking to all code snippets, including: 1) removing annotations directly referencing vulnerabilities; 2) replacing vulnerability-specific function names with generic ones (e.g., “Sample\_func()”); and 3) swapping variable names or tokens that suggest vulnerability (e.g., “good” or “bad”) with neutral terms such as “cat” or “apple.”

% \begin{figure}[t]
%     \centering
% \includegraphics[width=0.48\textwidth]{Figures/statistics.pdf}
%     \caption{Overview of benchmark question distribution across two core capabilities, five question types, and multiple difficulty levels. The notation \textit{-\textbackslash w CI} indicates code variants with control flow injection. The term \textit{None} indicates that there is no connection within the dataflow or control flow graph in the structure reasoning scenario. The question generation strategy is designed to ensure a uniform distribution across each base file, prioritizing high difficulty levels.}
%     \label{fig:data_statistic}
% \end{figure}

\begin{figure*}[t]
    \centering
    \includegraphics[width=\textwidth]{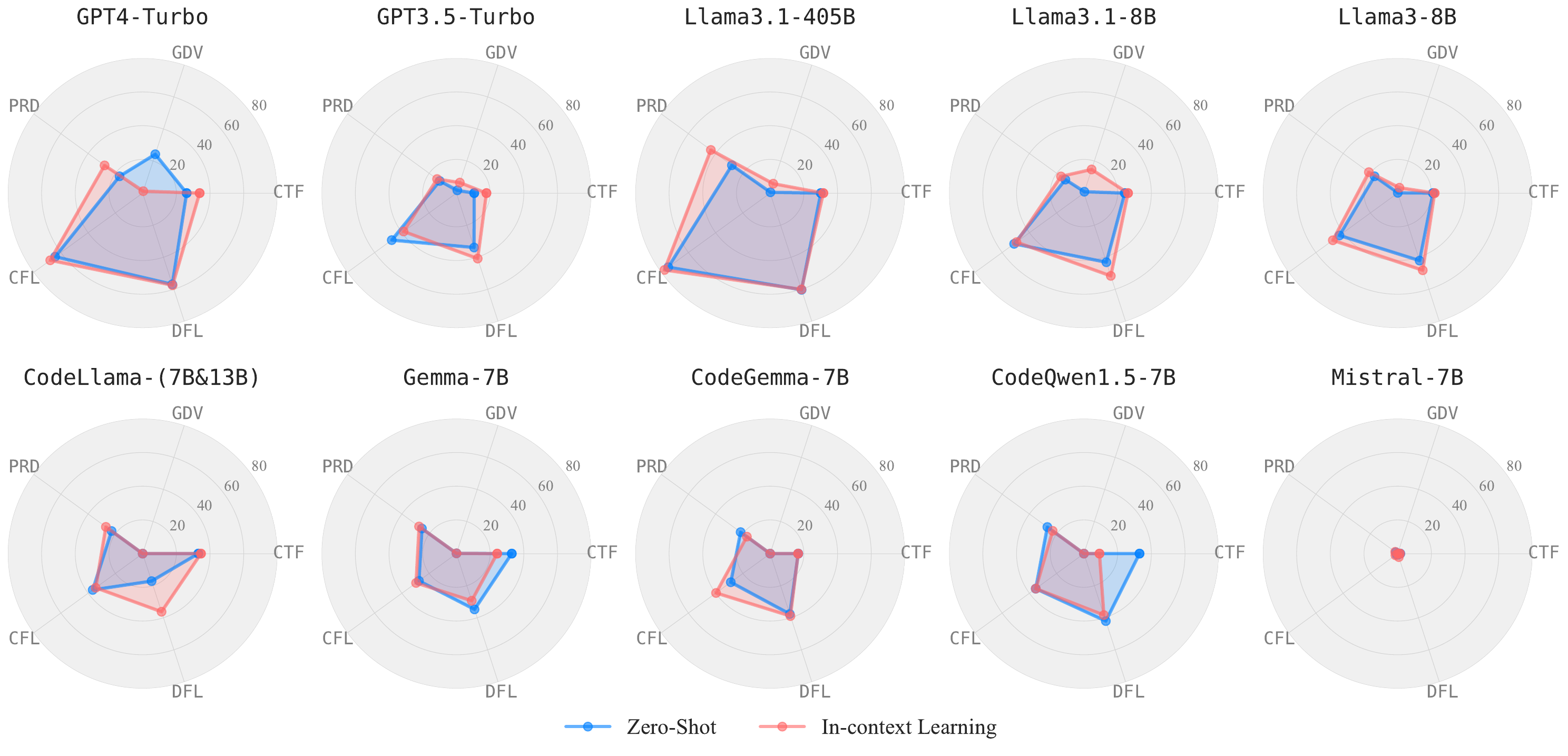}
    \caption{Overall consistency scores of LLMs on \textsc{SV-TrustEval-C}. For \texttt{CodeLlama}, the average consistency score across the 7B and 13B versions is reported due to their high similarity in performance. Abbreviations: \textbf{DFL} (DataFlow-wise questions), \textbf{CFL} (ControlFlow-wise questions), \textbf{CTF} (Counterfactual questions), \textbf{GDV} (Goal-driven questions), \textbf{PRD} (Predictive questions).}
    \label{fig:radar_plot}
\end{figure*}

\subsubsection{Question Design}
To develop an effective question-answering template that ensures clear comprehension by LLMs, we manually crafted seven distinct prompts for each question. We then employed GPT-4 as an automatic evaluator~\cite{zheng2024judging} to select the most suitable format for each QA. Additionally, to verify the absence of misleading syntax or semantics, we manually reviewed the intermediate explanations for 50 randomly selected QAs from each question type, ensuring that the LLMs’ responses were well-aligned with the QAs.

\subsection{Main Results}
As shown in Table~\ref{tab:main_res}, while LLMs generally require further improvement in structure reasoning, models with over 15 billion parameters significantly outperform their smaller counterparts. Notably, the recently released \texttt{Llama3.1B-405B} performs better in both data flow and control flow dimensions, achieving average scores of 68.58\% in zero-shot and 69.14\% in in-context learning, compared to \texttt{GPT-4}.

However, in semantic reasoning scenarios, most LLMs perform poorly, with average scores falling below 32\% for zero-shot and 46\% for in-context learning modes—particularly in Goal-driven tasks that require targeted vulnerability fixes and Predictive scenarios demanding extensive domain knowledge. We also conducted a baseline assessment by classifying code from the original Juliet Test Suite as \textit{safe} or \textit{unsafe} using a QA format with the options: A) Vulnerable, B) Non-Vulnerable, C) Do Not Know. The results presented in Table~\ref{tab:main_res} indicate that most models perform unsatisfactorily, especially given that the code originates from synthetically generated datasets like Juliet. This suggests that the models are inadequately pre-trained in code vulnerability detection.

Additionally, several models—including \texttt{CodeLlama}, \texttt{CodeQwen}, and \texttt{Gemma}—exhibit high false positive rates\footnote{For FPR$^{\text{safe}}$, false positives are instances where \textit{safe} code is incorrectly classified as \textit{unsafe} or as an unknown category; true negatives are cases where code is correctly classified as \textit{safe}.}, erroneously classifying \textit{safe} code as \textit{unsafe} in nearly 100\% of cases, as illustrated in Table~\ref{tab:main_res}. This raises concerns about the efficacy of LLMs in code analysis.

Finally, we observe that in-context learning can enhance models' performance in most cases. Specialized LLMs such as \texttt{CodeLlama} and \texttt{CodeGemma}, which are less effective at understanding complex natural language scenarios, benefit significantly from this approach. Conversely, general-purpose LLMs like \texttt{Mixtral}, which have less pre-training in the code domain, often struggle with code contexts; here, in-context learning serves as a bridge, helping these models adapt to intersecting scenarios. Similar improvements are also observed in models like \texttt{Llama3.1B-405B} and \texttt{GPT-3.5}. However, \texttt{GPT-4} exhibits a dramatic decline in performance in Goal-driven scenarios with in-context learning, indicating that different models may respond variably to this technique and may require specific customization for vulnerability analysis.

\subsubsection{Consistency Analysis}
Our consistency analysis assesses how models maintain reliable vulnerability evaluations across different scenarios by focusing on two distinct reasoning tasks: \textbf{structure reasoning} and \textbf{semantic reasoning}. In the structure reasoning scenario, we transition from base code vulnerability analysis to tasks that require understanding the code's structure, specifically DataFlow-wise (DFL) and ControlFlow-wise (CFL) analyses. Grasping the relationships among code components is essential for accurate vulnerability detection. If a model can identify vulnerabilities but fails to differentiate variable relationships within the source code, it indicates that its capabilities are driven by pre-trained patterns rather than genuine logical code analysis~\cite{chen2024reasoning,liu2024codemind}. Conversely, in the semantic reasoning scenario, we investigate whether LLMs can maintain consistent vulnerability assessments across diverse contexts by addressing the logical equivalence of statements such as $A = B$ and $B = A$~\cite{berglund2023reversal}. Using the base scenario as a benchmark, we evaluate if consistency is preserved in corresponding Counterfactual (CTF), Goal-driven (GDV), and Predictive (PRD) scenarios. This means that if LLMs accurately assess vulnerabilities in the base scenario, they should provide reliable analyses for any code variants within the same semantic context. For each question derived from the base scenario, we define the consistency scores as:

\begin{align}
    \text{Cons}_{\text{DFL}} &= \frac{ \sum_{i=1}^{N_{\text{DFL}}} \mathbb{I}\left(C_{\text{base}}^i = 1 \land C_{\text{DFL}}^i = 1\right) }{ N_{\text{DFL}} } \\
        \text{Cons}_{\text{CFL}} &= \frac{ \sum_{i=1}^{N_{\text{CFL}}} \mathbb{I}\left(C_{\text{base}}^i = 1 \land C_{\text{CFL}}^i = 1\right) }{ N_{\text{CFL}} } \\
            \text{Cons}_{\text{CTF}} &= \frac{ \sum_{i=1}^{N_{\text{CTF}}} \mathbb{I}\left(C_{\text{CTF}}^i = 1 \land C_{\text{base}}^i = 1\right) }{ N_{\text{CTF}} }\label{eq:CTF1}\\
    \text{Cons}_{\text{GDV}} &= \frac{ \sum_{i=1}^{N_{\text{GDV}}} \mathbb{I}\left(C_{\text{GDV}}^i = 1 \land C_{\text{safe}}^i = 1\right) }{ N_{\text{GDV}} } \\
    \text{Cons}_{\text{PRD}} &= \frac{ \sum_{i=1}^{N_{\text{PRD}}} \mathbb{I}\left(C_{\text{PRD}}^i = 1 \land C_{\text{unsafe}}^i = 1\right) }{ N_{\text{PRD}} }
\end{align}

\noindent where:
\begin{itemize}
    \item $\mathbb{I}(\cdot)$ is the indicator function, which equals 1 if the condition inside is true, and 0 otherwise.
     \item $C_{\text{safe}}^i$ and $C_{\text{unsafe}}^i$ denote the correctness indicators for \textit{safe} and \textit{unsafe} classifications in the base scenario for the $i$-th case.
    \item $C_{\text{base}}^i = \mathbb{I}\left( C_{\text{safe}}^i = 1 \land C_{\text{unsafe}}^i = 1 \right)$ is the correctness indicator for the base scenario for the \(i\)-th case.
    \item $C_{\text{DFL}}^i$ and $C_{\text{CFL}}^i$ are the correctness indicators for DataFlow-wise and ControlFlow-wise scenarios for the $i$-th case, respectively.
    \item $C_{\text{CTF}}^i$ is the correctness indicator for the Counterfactual scenario of the $i$-th case.
    \item $C_{\text{GDV}}^i$ and $C_{\text{PRD}}^i$ are the correctness indicators for Goal-driven and Predictive scenarios for the $i$-th case.
    \item $N_{\text{DFL}}$, $N_{\text{CFL}}$, $N_{\text{CTF}}$, $N_{\text{GDV}}$, and $N_{\text{PRD}}$ represent the total number of cases in the DataFlow-wise, ControlFlow-wise, Counterfactual, Goal-driven, and Predictive scenarios, respectively.
\end{itemize}

\begin{table*}[ht]
\centering
\caption{LLMs' performance across difficulty levels by question type. The symbol \faBan\ represents unknown parameter scales; terms \textbf{D-Flow} and \textbf{C-Flow} refer to DataFlow-wise and ControlFlow-wise questions, respectively; \textbf{O.\&I.} signifies the Outer\&Inner structure, \textbf{\#} denotes the number of questions, and \textit{CI} indicates code variants with control flow injection. ``None'' indicates that there are no direct or indirect connections between the target code elements within the code graphs. The \textcolor{crimsonred}{$\downarrow$} symbol denotes a decrease in model performance on the current task compared to \texttt{Zero} after introducing \texttt{ICL}, while \textcolor{deepskyblue}{$\uparrow$} signifies an improvement, and \textcolor{blue}{$\updownarrow$} indicates no change in performance.}
\resizebox{\textwidth}{!}{
\begin{tabular}[width=\textwidth]{cl|cccccccccccccccccccccc|l}
\toprule
&\textbf{LLMs} & \multicolumn{2}{c}{\textbf{\texttt{GPT-4}}} & \multicolumn{2}{c}{\textbf{\texttt{GPT-3.5}}} & \multicolumn{2}{c}{\textbf{\texttt{Llama3.1}}} & \multicolumn{2}{c}{\textbf{\texttt{Llama3.1}}} & \multicolumn{2}{c}{\textbf{\texttt{Llama3}}} & \multicolumn{2}{c}{\textbf{\texttt{CodeLlama}}} &
\multicolumn{2}{c}{\textbf{\texttt{CodeLlama}}} & \multicolumn{2}{c}{\textbf{\texttt{Gemma}}} & \multicolumn{2}{c}{\textbf{\texttt{CodeGemma}}} & \multicolumn{2}{c}{\textbf{\texttt{CodeQwen1.5}}} & \multicolumn{2}{c|}{\textbf{\texttt{Mixtral}}} &\multirow{3}{*}{\textbf{\#}}\\

&\textbf{Param.}& \multicolumn{2}{c}{\faBan} & \multicolumn{2}{c}{\faBan} & \multicolumn{2}{c}{\textbf{405B}} & \multicolumn{2}{c}{\textbf{8B}} & \multicolumn{2}{c}{\textbf{8B}} & \multicolumn{2}{c}{\textbf{13B}} & \multicolumn{2}{c}{\textbf{7B}}  & \multicolumn{2}{c}{\textbf{7B}} & \multicolumn{2}{c}{\textbf{7B}} & \multicolumn{2}{c}{\textbf{7B}} & \multicolumn{2}{c|}{\textbf{7B}} \\
&\textbf{Mode}& \texttt{Zero} & \texttt{ICL} & \texttt{Zero} & \texttt{ICL} & \texttt{Zero} & \texttt{ICL} & \texttt{Zero} & \texttt{ICL} & \texttt{Zero} & \texttt{ICL} & \texttt{Zero} & \texttt{ICL} & \texttt{Zero} & \texttt{ICL} &\texttt{Zero} & \texttt{ICL} & \texttt{Zero} & \texttt{ICL} & \texttt{Zero} & \texttt{ICL} & \texttt{Zero} & \texttt{ICL} \\
\midrule
\multicolumn{25}{c}{\cellcolor{gray!13}\textbf{Structure Reasoning}} \\
\midrule
\multirow{5}{*}{\rotatebox[origin=c]{90}{\textbf{D-Flow}}} & \textbf{\textit{None}}&81.66 & 82.04\textcolor{deepskyblue}{$\uparrow$} & 8.03 & 51.23\textcolor{deepskyblue}{$\uparrow$} & 81.76 & 80.91\textcolor{crimsonred}{$\downarrow$} & \cellcolor{lightgreencustom}{86.48} & 77.32\textcolor{crimsonred}{$\downarrow$} & 51.89 & \cellcolor{darkgreen!50}{\textcolor{white}{87.81}}\textcolor{deepskyblue}{$\uparrow$} & 2.08 & 0.00\textcolor{crimsonred}{$\downarrow$} & 4.63 & 26.47\textcolor{deepskyblue}{$\uparrow$} & 1.70 & 2.74\textcolor{deepskyblue}{$\uparrow$} & 0.00 & 1.89\textcolor{deepskyblue}{$\uparrow$} & 14.56 & 1.23\textcolor{crimsonred}{$\downarrow$} & 5.58 & 75.99\textcolor{deepskyblue}{$\uparrow$} & 1058\\

 & \textbf{1-Hop}&52.92 & 54.83\textcolor{deepskyblue}{$\uparrow$} & 72.71 & 62.74\textcolor{crimsonred}{$\downarrow$} & 57.85 & 59.57\textcolor{deepskyblue}{$\uparrow$} & 23.01 & 47.73\textcolor{deepskyblue}{$\uparrow$} & 50.91 & 30.92\textcolor{crimsonred}{$\downarrow$} & 70.69 & \cellcolor{lightgreencustom}{74.02}\textcolor{deepskyblue}{$\uparrow$} & 31.42 & 53.32\textcolor{deepskyblue}{$\uparrow$} & 68.08 & 57.60\textcolor{crimsonred}{$\downarrow$} & 73.72 & \cellcolor{darkgreen!50}{\textcolor{white}{74.92}}\textcolor{deepskyblue}{$\uparrow$} & 69.13 & 73.82\textcolor{deepskyblue}{$\uparrow$} & 40.38 & 38.32\textcolor{crimsonred}{$\downarrow$} & 993\\

 & \textbf{2-Hop}&32.25 & 26.87\textcolor{crimsonred}{$\downarrow$} & 74.43 & 39.09\textcolor{crimsonred}{$\downarrow$} & 36.32 & 35.67\textcolor{crimsonred}{$\downarrow$} & 11.89 & 33.22\textcolor{deepskyblue}{$\uparrow$} & 23.13 & 17.26\textcolor{crimsonred}{$\downarrow$} & 74.10 & \cellcolor{darkgreen!50}{\textcolor{white}{76.38}}\textcolor{deepskyblue}{$\uparrow$} & 46.09 & 44.79\textcolor{crimsonred}{$\downarrow$} & 64.33 & 53.91\textcolor{crimsonred}{$\downarrow$} & \cellcolor{lightgreencustom}{76.06} & 75.90\textcolor{crimsonred}{$\downarrow$} & 73.94 & \cellcolor{darkgreen!50}{\textcolor{white}{76.38}}\textcolor{deepskyblue}{$\uparrow$} & 40.23 & 32.25\textcolor{crimsonred}{$\downarrow$} & 307\\
 
 & \textbf{3-Hop}&51.61 & 50.00\textcolor{crimsonred}{$\downarrow$} & \cellcolor{lightgreencustom}{79.03} & 67.74\textcolor{crimsonred}{$\downarrow$} & 53.23 & 45.16\textcolor{crimsonred}{$\downarrow$} & 34.68 & 49.19\textcolor{deepskyblue}{$\uparrow$} & 58.06 & 33.87\textcolor{crimsonred}{$\downarrow$} & 70.16 & \cellcolor{darkgreen!50}{\textcolor{white}{79.84}}\textcolor{deepskyblue}{$\uparrow$} & 42.74 & 70.97\textcolor{deepskyblue}{$\uparrow$} & \cellcolor{darkgreen!50}{\textcolor{white}{79.84}} & 67.74\textcolor{crimsonred}{$\downarrow$} & \cellcolor{darkgreen!50}{\textcolor{white}{79.84}} & 78.23\textcolor{crimsonred}{$\downarrow$} & \cellcolor{lightgreencustom}{79.03} & \cellcolor{darkgreen!50}{\textcolor{white}{79.84}}\textcolor{deepskyblue}{$\uparrow$} & 47.58 & 54.84\textcolor{deepskyblue}{$\uparrow$} & 62\\
 
 & \textbf{4-Hop}&30.00 & 30.00\textcolor{blue}{$\updownarrow$} & \cellcolor{darkgreen!50}{\textcolor{white}{100.00}} & 80.00\textcolor{crimsonred}{$\downarrow$} & 0.00 & 10.00\textcolor{deepskyblue}{$\uparrow$} & 40.00 & 70.00\textcolor{deepskyblue}{$\uparrow$} & 30.00 & 0.00\textcolor{crimsonred}{$\downarrow$} & \cellcolor{lightgreencustom}{90.00} & \cellcolor{darkgreen!50}{\textcolor{white}{100.00}}\textcolor{deepskyblue}{$\uparrow$} & 0.00 & \cellcolor{darkgreen!50}{\textcolor{white}{100.00}}\textcolor{deepskyblue}{$\uparrow$} & \cellcolor{darkgreen!50}{\textcolor{white}{100.00}} & \cellcolor{darkgreen!50}{\textcolor{white}{100.00}}\textcolor{blue}{$\updownarrow$} & \cellcolor{darkgreen!50}{\textcolor{white}{100.00}} & \cellcolor{darkgreen!50}{\textcolor{white}{100.00}}\textcolor{blue}{$\updownarrow$} & \cellcolor{darkgreen!50}{\textcolor{white}{100.00}} & \cellcolor{darkgreen!50}{\textcolor{white}{100.00}}\textcolor{blue}{$\updownarrow$} & 0.00 & 30.00\textcolor{deepskyblue}{$\uparrow$} & 10\\
\midrule
\multirow{4}{*}{\rotatebox[origin=c]{90}{\textbf{C-Flow}}} & \textbf{\textit{None}}&73.77 & 80.52\textcolor{deepskyblue}{$\uparrow$} & 52.76 & 49.23\textcolor{crimsonred}{$\downarrow$} & 87.27 & 82.67\textcolor{crimsonred}{$\downarrow$} & 90.80 & \cellcolor{darkgreen!50}{\textcolor{white}{94.63}}\textcolor{deepskyblue}{$\uparrow$} & 71.32 & \cellcolor{lightgreencustom}{93.71}\textcolor{deepskyblue}{$\uparrow$} & 2.30 & 0.92\textcolor{crimsonred}{$\downarrow$} & 3.07 & 8.74\textcolor{deepskyblue}{$\uparrow$} & 0.00 & 0.00\textcolor{blue}{$\updownarrow$} & 0.15 & 14.72\textcolor{deepskyblue}{$\uparrow$} & 0.15 & 0.77\textcolor{deepskyblue}{$\uparrow$} & 28.07 & 70.86\textcolor{deepskyblue}{$\uparrow$} & 652\\
 & \textbf{1-Hop}&81.39 & 81.84\textcolor{deepskyblue}{$\uparrow$} & 83.18 & 61.21\textcolor{crimsonred}{$\downarrow$} & 87.67 & 88.12\textcolor{deepskyblue}{$\uparrow$} & 29.37 & 18.83\textcolor{crimsonred}{$\downarrow$} & 19.73 & 5.38\textcolor{crimsonred}{$\downarrow$} & 95.96 & 79.37\textcolor{crimsonred}{$\downarrow$} & 97.31 & 80.72\textcolor{crimsonred}{$\downarrow$} & 30.04 & 42.38\textcolor{deepskyblue}{$\uparrow$} & 49.33 & 46.86\textcolor{crimsonred}{$\downarrow$} & \cellcolor{darkgreen!50}{\textcolor{white}{100.00}} & \cellcolor{lightgreencustom}{98.88}\textcolor{crimsonred}{$\downarrow$} & 40.13 & 43.05\textcolor{deepskyblue}{$\uparrow$} & 446\\
 & \textbf{2-Hop}&16.11 & 13.89\textcolor{crimsonred}{$\downarrow$} & 24.44 & 20.00\textcolor{crimsonred}{$\downarrow$} & 14.44 & 32.22\textcolor{deepskyblue}{$\uparrow$} & 7.78 & 1.67\textcolor{crimsonred}{$\downarrow$} & 22.78 & 1.67\textcolor{crimsonred}{$\downarrow$} & 12.78 & 15.56\textcolor{deepskyblue}{$\uparrow$} & 20.00 & 15.00\textcolor{crimsonred}{$\downarrow$} & \cellcolor{lightgreencustom}{96.67} & 92.22\textcolor{crimsonred}{$\downarrow$} & 67.78 & \cellcolor{darkgreen!50}{\textcolor{white}{98.89}}\textcolor{deepskyblue}{$\uparrow$} & 15.56 & 15.56\textcolor{blue}{$\updownarrow$} & 63.33 & 15.00\textcolor{crimsonred}{$\downarrow$} & 180\\
 
 & \textbf{3-Hop}&5.97 & 10.45\textcolor{deepskyblue}{$\uparrow$} & 10.45 & 10.45\textcolor{blue}{$\updownarrow$} & 13.43 & 65.67\textcolor{deepskyblue}{$\uparrow$} & 2.99 & 5.97\textcolor{deepskyblue}{$\uparrow$} & 10.45 & 20.90\textcolor{deepskyblue}{$\uparrow$} & 0.00 & 7.46\textcolor{deepskyblue}{$\uparrow$} & 0.00 & 26.87\textcolor{deepskyblue}{$\uparrow$} & \cellcolor{darkgreen!50}{\textcolor{white}{100.00}} & 71.64\textcolor{crimsonred}{$\downarrow$} & 76.12 & \cellcolor{lightgreencustom}{79.10}\textcolor{deepskyblue}{$\uparrow$} & 0.00 & 0.00\textcolor{blue}{$\updownarrow$} & 46.27 & 37.31\textcolor{crimsonred}{$\downarrow$} & 67\\
 
\midrule
\multicolumn{25}{c}{\cellcolor{gray!13}\textbf{Semantic Reasoning}} \\
\midrule
\multirow{6}{*}{\rotatebox[origin=c]{90}{\textbf{Counterfactual}}} & \textbf{Outer} & 43.59 & \cellcolor{lightgreencustom}{58.00}\textcolor{deepskyblue}{$\uparrow$} & 35.46 & 34.84\textcolor{crimsonred}{$\downarrow$} & 51.37 & \cellcolor{darkgreen!50}{\textcolor{white}{68.52}}\textcolor{deepskyblue}{$\uparrow$} & 37.22 & 38.11\textcolor{deepskyblue}{$\uparrow$} & 32.71 & 34.31\textcolor{deepskyblue}{$\uparrow$} & 33.24 & 33.33\textcolor{deepskyblue}{$\uparrow$} & 33.33 & 33.33\textcolor{blue}{$\updownarrow$} & 32.54 & 35.37\textcolor{deepskyblue}{$\uparrow$} & 33.42 & 34.66\textcolor{deepskyblue}{$\uparrow$} & 32.54 & 31.56\textcolor{crimsonred}{$\downarrow$} & 33.51 & 36.07\textcolor{deepskyblue}{$\uparrow$} & 1131\\

 & \quad-\textbackslash\textit{w} \textit{CI}&41.46 & \cellcolor{lightgreencustom}{50.00}\textcolor{deepskyblue}{$\uparrow$} & 35.77 & 32.32\textcolor{crimsonred}{$\downarrow$} & 38.62 & \cellcolor{darkgreen!50}{\textcolor{white}{55.28}}\textcolor{deepskyblue}{$\uparrow$} & 36.99 & 34.15\textcolor{crimsonred}{$\downarrow$} & 31.10 & 32.32\textcolor{deepskyblue}{$\uparrow$} & 32.32 & 32.93\textcolor{deepskyblue}{$\uparrow$} & 33.13 & 32.93\textcolor{crimsonred}{$\downarrow$} & 32.93 & 34.55\textcolor{deepskyblue}{$\uparrow$} & 35.57 & 32.72\textcolor{crimsonred}{$\downarrow$} & 32.93 & 32.52\textcolor{crimsonred}{$\downarrow$} & 36.59 & 31.91\textcolor{crimsonred}{$\downarrow$} & 492\\
 
 & \textbf{Inner}&45.73 & \cellcolor{lightgreencustom}{53.38}\textcolor{deepskyblue}{$\uparrow$} & 38.79 & 40.04\textcolor{deepskyblue}{$\uparrow$} & 44.48 & \cellcolor{darkgreen!50}{\textcolor{white}{62.81}}\textcolor{deepskyblue}{$\uparrow$} & 40.04 & 39.32\textcolor{crimsonred}{$\downarrow$} & 37.19 & 37.90\textcolor{deepskyblue}{$\uparrow$} & 37.54 & 37.54\textcolor{blue}{$\updownarrow$} & 37.54 & 37.54\textcolor{blue}{$\updownarrow$} & 32.38 & 31.67\textcolor{crimsonred}{$\downarrow$} & 38.43 & 37.72\textcolor{crimsonred}{$\downarrow$} & 32.38 & 36.65\textcolor{deepskyblue}{$\uparrow$} & 37.72 & 37.19\textcolor{crimsonred}{$\downarrow$} & 562\\

  & \quad-\textbackslash\textit{w} \textit{CI}&42.69 & \cellcolor{lightgreencustom}{48.72}\textcolor{deepskyblue}{$\uparrow$} & 37.59 & 39.68\textcolor{deepskyblue}{$\uparrow$} & 46.87 & \cellcolor{darkgreen!50}{\textcolor{white}{61.95}}\textcolor{deepskyblue}{$\uparrow$} & 38.28 & 42.92\textcolor{deepskyblue}{$\uparrow$} & 39.21 & 39.21\textcolor{blue}{$\updownarrow$} & 39.21 & 39.21\textcolor{blue}{$\updownarrow$} & 38.75 & 39.21\textcolor{deepskyblue}{$\uparrow$} & 37.12 & 32.02\textcolor{crimsonred}{$\downarrow$} & 36.89 & 38.52\textcolor{deepskyblue}{$\uparrow$} & 37.12 & 33.41\textcolor{crimsonred}{$\downarrow$} & 37.35 & 34.11\textcolor{crimsonred}{$\downarrow$} & 431\\
  
 & \textbf{O.\&I.}&36.77 & \cellcolor{lightgreencustom}{50.87}\textcolor{deepskyblue}{$\uparrow$} & 35.82 & 32.65\textcolor{crimsonred}{$\downarrow$} & 43.58 & \cellcolor{darkgreen!50}{\textcolor{white}{62.92}}\textcolor{deepskyblue}{$\uparrow$} & 34.07 & 38.19\textcolor{deepskyblue}{$\uparrow$} & 36.61 & 35.18\textcolor{crimsonred}{$\downarrow$} & 33.91 & 33.44\textcolor{crimsonred}{$\downarrow$} & 33.28 & 33.44\textcolor{deepskyblue}{$\uparrow$} & 32.49 & 33.28\textcolor{deepskyblue}{$\uparrow$} & 32.96 & 32.17\textcolor{crimsonred}{$\downarrow$} & 32.49 & 32.33\textcolor{crimsonred}{$\downarrow$} & 35.66 & 28.05\textcolor{crimsonred}{$\downarrow$} & 631\\
 
  & \quad-\textbackslash\textit{w} \textit{CI}&37.52 & \cellcolor{lightgreencustom}{45.11}\textcolor{deepskyblue}{$\uparrow$} & 32.93 & 32.53\textcolor{crimsonred}{$\downarrow$} & 38.92 & \cellcolor{darkgreen!50}{\textcolor{white}{56.09}}\textcolor{deepskyblue}{$\uparrow$} & 34.73 & 38.72\textcolor{deepskyblue}{$\uparrow$} & 34.53 & 34.53\textcolor{blue}{$\updownarrow$} & 32.93 & 33.33\textcolor{deepskyblue}{$\uparrow$} & 32.73 & 33.33\textcolor{deepskyblue}{$\uparrow$} & 32.73 & 32.73\textcolor{blue}{$\updownarrow$} & 35.13 & 36.73\textcolor{deepskyblue}{$\uparrow$} & 32.73 & 31.94\textcolor{crimsonred}{$\downarrow$} & 35.73 & 28.14\textcolor{crimsonred}{$\downarrow$} & 501\\
 \midrule
 
\multirow{6}{*}{\rotatebox[origin=c]{90}{\textbf{Goal-driven}}} & \textbf{Outer} & \cellcolor{darkgreen!50}{\textcolor{white}{66.31}} & 2.39\textcolor{crimsonred}{$\downarrow$} & 1.33 & 15.92\textcolor{deepskyblue}{$\uparrow$} & 2.39 & 27.59\textcolor{deepskyblue}{$\uparrow$} & 0.00 & 24.67\textcolor{deepskyblue}{$\uparrow$} & 0.00 & 7.43\textcolor{deepskyblue}{$\uparrow$} & 0.00 & 34.22\textcolor{deepskyblue}{$\uparrow$} & 4.51 & 20.69\textcolor{deepskyblue}{$\uparrow$} & 5.31 & \cellcolor{lightgreencustom}{35.54}\textcolor{deepskyblue}{$\uparrow$} & 0.00 & 23.34\textcolor{deepskyblue}{$\uparrow$} & 34.22 & 22.55\textcolor{crimsonred}{$\downarrow$} & 0.00 & 8.49\textcolor{deepskyblue}{$\uparrow$} & 377\\

 & \quad-\textbackslash\textit{w} \textit{CI} & 0.61 & 0.00\textcolor{crimsonred}{$\downarrow$} & 21.21 & 19.39\textcolor{crimsonred}{$\downarrow$} & 7.88 & 37.58\textcolor{deepskyblue}{$\uparrow$} & 3.64 & 22.42\textcolor{deepskyblue}{$\uparrow$} & 0.00 & 4.24\textcolor{deepskyblue}{$\uparrow$} & 0.00 & 29.09\textcolor{deepskyblue}{$\uparrow$} & \cellcolor{darkgreen!50}{\textcolor{white}{54.55}} & 24.85\textcolor{crimsonred}{$\downarrow$} & \cellcolor{lightgreencustom}{46.06} & 42.42\textcolor{crimsonred}{$\downarrow$} & 0.00 & 27.27\textcolor{deepskyblue}{$\uparrow$} & 36.36 & 27.88\textcolor{crimsonred}{$\downarrow$} & 0.00 & 16.36\textcolor{deepskyblue}{$\uparrow$} & 165\\

 & \textbf{Inner} & \cellcolor{darkgreen!50}{\textcolor{white}{55.56}} & 2.22\textcolor{crimsonred}{$\downarrow$} & 2.22 & 17.04\textcolor{deepskyblue}{$\uparrow$} & 0.00 & 18.52\textcolor{deepskyblue}{$\uparrow$} & 0.00 & 23.70\textcolor{deepskyblue}{$\uparrow$} & 0.00 & 8.15\textcolor{deepskyblue}{$\uparrow$} & 0.00 & 39.26\textcolor{deepskyblue}{$\uparrow$} & 11.11 & 22.22\textcolor{deepskyblue}{$\uparrow$} & 7.41 & \cellcolor{lightgreencustom}{45.93}\textcolor{deepskyblue}{$\uparrow$} & 0.00 & 20.74\textcolor{deepskyblue}{$\uparrow$} & 25.19 & 23.70\textcolor{crimsonred}{$\downarrow$} & 0.00 & 7.41\textcolor{deepskyblue}{$\uparrow$} & 135\\
 
  & \quad-\textbackslash\textit{w} \textit{CI} & 4.76 & 1.90\textcolor{crimsonred}{$\downarrow$} & 1.90 & 18.10\textcolor{deepskyblue}{$\uparrow$} & 0.00 & 33.33\textcolor{deepskyblue}{$\uparrow$} & 3.81 & 26.67\textcolor{deepskyblue}{$\uparrow$} & 10.48 & 1.90\textcolor{crimsonred}{$\downarrow$} & 0.00 & 31.43\textcolor{deepskyblue}{$\uparrow$} & \cellcolor{darkgreen!50}{\textcolor{white}{41.90}} & 24.76\textcolor{crimsonred}{$\downarrow$} & 30.48 & \cellcolor{darkgreen!50}{\textcolor{white}{41.90}}\textcolor{deepskyblue}{$\uparrow$} & 0.00 & 22.86\textcolor{deepskyblue}{$\uparrow$} & \cellcolor{lightgreencustom}{34.29} & 25.71\textcolor{crimsonred}{$\downarrow$} & 0.00 & 3.81\textcolor{deepskyblue}{$\uparrow$} & 105\\
  
 & \textbf{O.\&I.} & 13.81 & 1.90\textcolor{crimsonred}{$\downarrow$} & 1.43 & 18.57\textcolor{deepskyblue}{$\uparrow$} & 1.43 & 30.00\textcolor{deepskyblue}{$\uparrow$} & 0.00 & 18.57\textcolor{deepskyblue}{$\uparrow$} & 0.00 & 4.76\textcolor{deepskyblue}{$\uparrow$} & 0.00 & \cellcolor{lightgreencustom}{34.29}\textcolor{deepskyblue}{$\uparrow$} & 21.43 & 23.81\textcolor{deepskyblue}{$\uparrow$} & 33.81 & \cellcolor{darkgreen!50}{\textcolor{white}{47.62}}\textcolor{deepskyblue}{$\uparrow$} & 0.00 & 21.90\textcolor{deepskyblue}{$\uparrow$} & 32.38 & 15.71\textcolor{crimsonred}{$\downarrow$} & 0.00 & 7.62\textcolor{deepskyblue}{$\uparrow$} & 210\\
 
  & \quad-\textbackslash\textit{w} \textit{CI} &1.20 & 0.00\textcolor{crimsonred}{$\downarrow$} & 4.79 & 15.57\textcolor{deepskyblue}{$\uparrow$} & 3.59 & 28.74\textcolor{deepskyblue}{$\uparrow$} & 1.80 & 22.75\textcolor{deepskyblue}{$\uparrow$} & 0.00 & 7.19\textcolor{deepskyblue}{$\uparrow$} & 0.00 & 33.53\textcolor{deepskyblue}{$\uparrow$} & \cellcolor{lightgreencustom}{44.91} & 27.54\textcolor{crimsonred}{$\downarrow$} & \cellcolor{darkgreen!50}{\textcolor{white}{75.45}} & 43.11\textcolor{crimsonred}{$\downarrow$} & 0.60 & 19.76\textcolor{deepskyblue}{$\uparrow$} & 38.32 & 23.95\textcolor{crimsonred}{$\downarrow$} & 0.00 & 13.77\textcolor{deepskyblue}{$\uparrow$} & 167\\
 \midrule
 
\multirow{6}{*}{\rotatebox[origin=c]{90}{\textbf{Predictive}}} & \textbf{Outer} & 19.37 & 25.23\textcolor{deepskyblue}{$\uparrow$} & 16.22 & 25.68\textcolor{deepskyblue}{$\uparrow$} & 27.93 & \cellcolor{darkgreen!50}{\textcolor{white}{38.74}}\textcolor{deepskyblue}{$\uparrow$} & 31.08 & 28.38\textcolor{crimsonred}{$\downarrow$} & 18.02 & 27.48\textcolor{deepskyblue}{$\uparrow$} & 23.87 & 31.08\textcolor{deepskyblue}{$\uparrow$} & 23.42 & 28.38\textcolor{deepskyblue}{$\uparrow$} & \cellcolor{lightgreencustom}{32.88} & 25.23\textcolor{crimsonred}{$\downarrow$} & 24.32 & 9.46\textcolor{crimsonred}{$\downarrow$} & 27.03 & 24.32\textcolor{crimsonred}{$\downarrow$} & 31.53 & 27.48\textcolor{crimsonred}{$\downarrow$} & 222\\

 & \quad-\textbackslash\textit{w} \textit{CI} & 15.89 & 29.91\textcolor{deepskyblue}{$\uparrow$} & 18.69 & 24.30\textcolor{deepskyblue}{$\uparrow$} & 26.17 & \cellcolor{darkgreen!50}{\textcolor{white}{48.60}}\textcolor{deepskyblue}{$\uparrow$} & 18.69 & 25.23\textcolor{deepskyblue}{$\uparrow$} & 28.97 & 28.97\textcolor{blue}{$\updownarrow$} & 23.36 & 23.36\textcolor{blue}{$\updownarrow$} & 26.17 & 22.43\textcolor{crimsonred}{$\downarrow$} & 17.76 & 31.78\textcolor{deepskyblue}{$\uparrow$} & 16.82 & 24.30\textcolor{deepskyblue}{$\uparrow$} & 28.04 & 17.76\textcolor{crimsonred}{$\downarrow$} & 30.84 & \cellcolor{lightgreencustom}{39.25}\textcolor{deepskyblue}{$\uparrow$} & 107\\

 & \textbf{Inner}&26.88 & 34.41\textcolor{deepskyblue}{$\uparrow$} & 23.66 & 15.05\textcolor{crimsonred}{$\downarrow$} & 32.26 & \cellcolor{darkgreen!50}{\textcolor{white}{47.31}}\textcolor{deepskyblue}{$\uparrow$} & 19.35 & 25.81\textcolor{deepskyblue}{$\uparrow$} & 22.58 & 24.73\textcolor{deepskyblue}{$\uparrow$} & 31.18 & 29.03\textcolor{crimsonred}{$\downarrow$} & 22.58 & 32.26\textcolor{deepskyblue}{$\uparrow$} & 18.28 & 27.96\textcolor{deepskyblue}{$\uparrow$} & 26.88 & 17.20\textcolor{crimsonred}{$\downarrow$} & 24.73 & 25.81\textcolor{deepskyblue}{$\uparrow$} & 24.73 & \cellcolor{lightgreencustom}{35.48}\textcolor{deepskyblue}{$\uparrow$} & 93\\

  & \quad-\textbackslash\textit{w} \textit{CI}&25.40 & 34.92\textcolor{deepskyblue}{$\uparrow$} & 22.22 & 22.22\textcolor{blue}{$\updownarrow$} & 30.16 & \cellcolor{darkgreen!50}{\textcolor{white}{42.86}}\textcolor{deepskyblue}{$\uparrow$} & 23.81 & 26.98\textcolor{deepskyblue}{$\uparrow$} & 26.98 & 34.92\textcolor{deepskyblue}{$\uparrow$} & 25.40 & 34.92\textcolor{deepskyblue}{$\uparrow$} & 30.16 & 25.40\textcolor{crimsonred}{$\downarrow$} & 34.92 & 30.16\textcolor{crimsonred}{$\downarrow$} & 23.81 & 20.63\textcolor{crimsonred}{$\downarrow$} & \cellcolor{lightgreencustom}{36.51} & 17.46\textcolor{crimsonred}{$\downarrow$} & 33.33 & 30.16\textcolor{crimsonred}{$\downarrow$} & 63\\

 & \textbf{O.\&I.}&19.12 & \cellcolor{lightgreencustom}{32.35}\textcolor{deepskyblue}{$\uparrow$} & 19.12 & 25.74\textcolor{deepskyblue}{$\uparrow$} & 30.88 & \cellcolor{darkgreen!50}{\textcolor{white}{46.32}}\textcolor{deepskyblue}{$\uparrow$} & 24.26 & 27.94\textcolor{deepskyblue}{$\uparrow$} & 26.47 & 28.68\textcolor{deepskyblue}{$\uparrow$} & 24.26 & 27.21\textcolor{deepskyblue}{$\uparrow$} & 17.65 & 30.88\textcolor{deepskyblue}{$\uparrow$} & 23.53 & 18.38\textcolor{crimsonred}{$\downarrow$} & 19.12 & 21.32\textcolor{deepskyblue}{$\uparrow$} & 25.74 & 22.79\textcolor{crimsonred}{$\downarrow$} & 27.94 & 27.94\textcolor{blue}{$\updownarrow$} & 136\\
 
  & \quad-\textbackslash\textit{w} \textit{CI} &14.29 & 33.67\textcolor{deepskyblue}{$\uparrow$} & 22.45 & 24.49\textcolor{deepskyblue}{$\uparrow$} & 27.55 & \cellcolor{darkgreen!50}{\textcolor{white}{44.90}}\textcolor{deepskyblue}{$\uparrow$} & 24.49 & 29.59\textcolor{deepskyblue}{$\uparrow$} & 27.55 & \cellcolor{lightgreencustom}{37.76}\textcolor{deepskyblue}{$\uparrow$} & 21.43 & 29.59\textcolor{deepskyblue}{$\uparrow$} & 20.41 & 19.39\textcolor{crimsonred}{$\downarrow$} & 19.39 & \cellcolor{lightgreencustom}{37.76}\textcolor{deepskyblue}{$\uparrow$} & 18.37 & 19.39\textcolor{deepskyblue}{$\uparrow$} & 22.45 & 26.53\textcolor{deepskyblue}{$\uparrow$} & 32.65 & \cellcolor{lightgreencustom}{37.76}\textcolor{deepskyblue}{$\uparrow$} & 98\\
\bottomrule
\end{tabular}}
\label{tab:diff_scale}
\end{table*}

\noindent
\textbf{Observations:} As shown in Figure~\ref{fig:radar_plot}, \texttt{Mistral-7B} exhibits an almost negligible consistency score in all scenarios, largely due to its weak performance in the Base scenarios. By contrast, \texttt{GPT-4} and \texttt{Llama3.1-405B} substantially outperform other models in both \(\text{Cons}_{\text{CFL}}\) and \(\text{Cons}_{\text{DFL}}\), reflecting a stronger ability to track dependencies between control statements, variables, and functions within the code. In comparison, the remaining LLMs achieve under 60\% consistency in structure reasoning scenarios, indicating that effectiveness in vulnerability analysis alone does not necessarily translate into a robust understanding of these code elements. In the semantic reasoning scenario, the consistency scores for all LLMs are notably low, averaging below 50\%. This suggests that, despite some LLMs being pre-trained in the code domain, they struggle to maintain consistent analysis across variant scenarios, even when these variants incorporate the original code snippet. Particularly in Goal-driven scenarios, all LLMs exhibit minimal consistency. For instance, \texttt{CodeQwen1.5} and \texttt{Gemma}, despite achieving the highest scores of 33.74\% and 41.59\% respectively in zero-shot and in-context learning modes for Goal-driven scenarios, record a 0\% $\text{Cons}_{\text{GDV}}$. This indicates that their success does not reflect a genuine understanding of code vulnerabilities. Similarly, models like \texttt{Llama3} and \texttt{Llama3.1} have nearly zero consistency scores, highlighting a failure to effectively apply their knowledge of vulnerabilities in practical scenarios. While introducing in-context learning generally improves $\text{Cons}_{\text{GDV}}$ and $\text{Cons}_{\text{PRD}}$, the enhancements are insufficient, and further efforts are required to bolster LLMs' consistency in semantic reasoning tasks. This underscores the necessity for more specialized training and fine-tuning to enable LLMs to better understand and analyze code vulnerabilities across diverse and complex scenarios.

\begin{figure*}
    \centering
    \includegraphics[width=\textwidth]{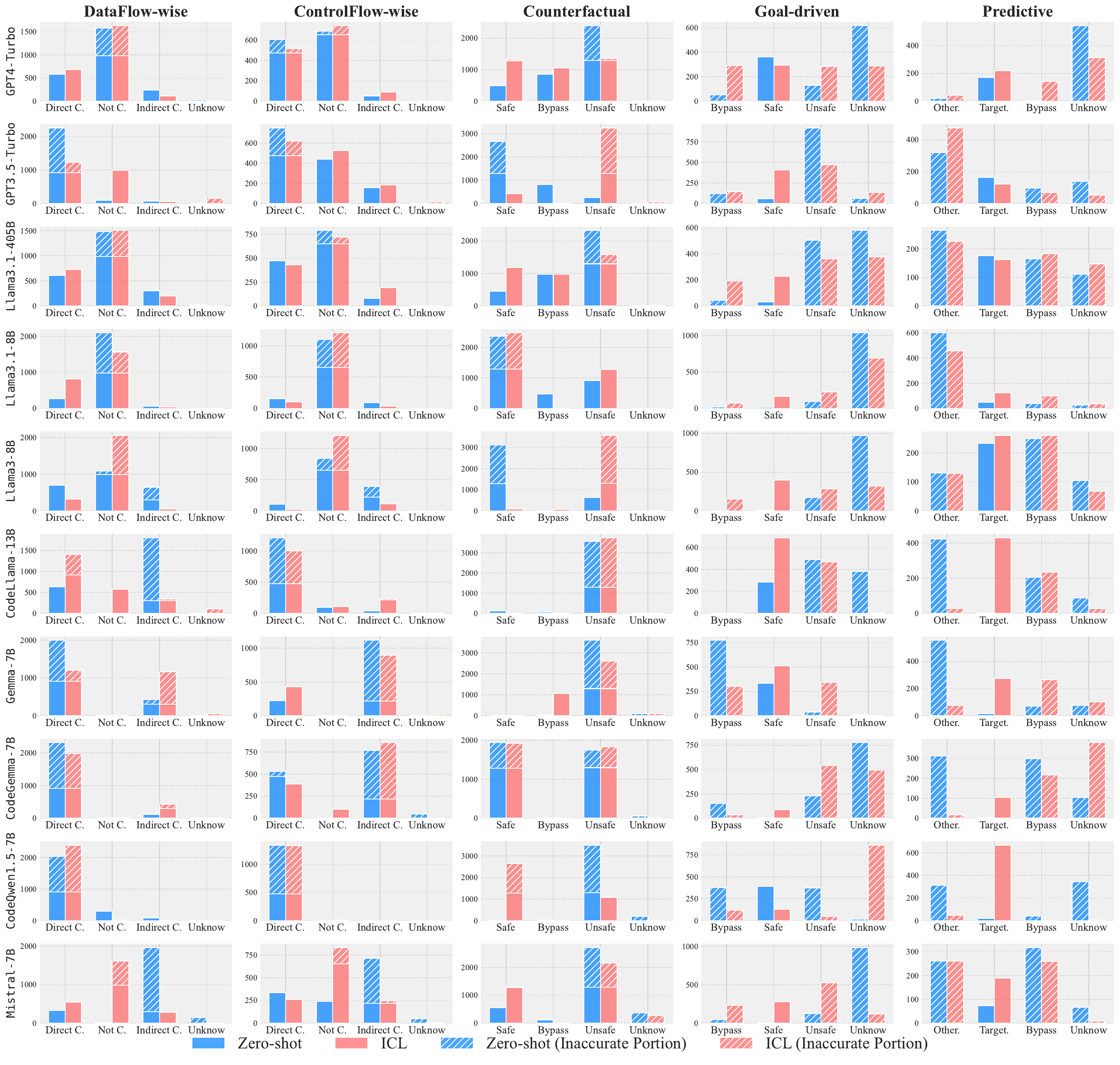}
    \caption{Distribution of choices made by all LLMs across all question scenarios, with the \textbf{shaded portions} of each bar showing how much the predicted option exceeds its corresponding ground truth.}
    \label{fig:choice_dirtribution}
\end{figure*}

\section{Analysis}
\subsection{Effects of Difficulty}
\label{sec:effects_of_difficulty}

As detailed in Table~\ref{tab:diff_scale}, we evaluated the performance of various LLMs across different difficulty levels within each scenario. The analysis encompasses three primary scenarios: Counterfactual, Goal-driven, and Predictive.

\subsubsection{Structure Reasoning}
In the structure reasoning tasks for DataFlow and ControlFlow scenarios, the LLMs' performance generally falls into two categories:

\noindent
\textbf{High Performance at the ``None'' Level, but Struggles with Complexity:} Models such as \texttt{Llama3.1} and \texttt{GPT-4} perform well at detecting when no connection exists between two code elements (the "None" level), but they face challenges with more complex relationships. However, their performance declines significantly in more complex scenarios that involve connections of three or more hops. This indicates a robust foundational understanding that diminishes as the reasoning required becomes more intricate.

\noindent
\textbf{Low Performance at the ``None" Level, but Improved with Increased Complexity:} Conversely, models like \texttt{CodeQwen}, \texttt{CodeLlama}, and \texttt{CodeGemma} exhibit average performance below 30\% at the ``\textit{None}'' level. Surprisingly, these models perform better in scenarios requiring connections of $\geq3$ hops. This counterintuitive outcome arises because these LLMs tend to assume connectivity among all code elements within a context, thereby increasing the likelihood of selecting correct answers. Specifically, explicit and implicit connections account for 50\% of the options (Options A and C), with some scenarios allowing both A and C to be simultaneously correct.

Despite these patterns, the ability to identify that two code elements are unconnected remains crucial, as it demonstrates the LLMs' capacity for logical and critical analysis of relationships between code elements. Introducing in-context learning significantly improved the ``\textit{None}'' level performance for structure reasoning tasks in most cases. For instance, \texttt{Mixtral} showed a dataflow-wise performance increase from 5.58\% to 75.99\%, and a controlflow-wise increase from 28.07\% to 70.86\%. However, this enhancement was not consistently observed across other difficulty levels for most LLMs. This inconsistency suggests that inherent code structure reasoning abilities require more sophisticated training methodologies and advanced model architectures to effectively handle varying levels of complexity.

\subsubsection{Vulnerability Reasoning}
Our evaluation across various scenarios reveals the following key insights:

\begin{itemize}
    \item \textbf{Model Size and Complexity Handling:} Larger models (e.g., \texttt{GPT-4}, \texttt{Llama3.1}) typically perform better in Counterfactual and Goal-driven tasks, especially with complex code variants. However, they show comparatively weaker performance on vulnerability-resolving tasks, suggesting a need for more advanced reasoning capabilities.
    
    \item \textbf{Impact of In-Context Learning:} In-context learning often leads to performance gains for models such as \texttt{CodeLlama}, \texttt{CodeGemma}, and \texttt{Llama3.1}-serious, particularly in Goal-driven scenarios. Conversely, models like \texttt{CodeQwen1.5} exhibit consistent declines, highlighting varying sensitivities to in-context prompts and emphasizing the need for tailored prompt adaptation. These mixed outcomes underscore that model-specific strategies are crucial.
    
    \item \textbf{Impact of Control Flow Injection:} Most LLMs experience performance degradation after control flow injection (-\textbackslash\textit{w} \textit{CI}), indicating that even minimal syntactic perturbations, without altering semantics, can adversely affect performance in diverse scenarios.
\end{itemize}

\noindent
\textbf{Counterfactual Scenarios.} Larger models such as \texttt{GPT-4} and \texttt{Llama3.1} consistently outperform smaller open-source counterparts (e.g., \texttt{CodeQwen}, \texttt{Mistral}), though their overall performance wanes as code complexity grows. In-context learning generally enhances the capabilities of \texttt{GPT-4} and \texttt{Llama3.1}, whereas it negatively impacts \texttt{CodeQwen} and \texttt{Mistral}. This discrepancy suggests that architectural differences or training data variations impact how models respond to additional contextual information.

\noindent
\textbf{Goal-driven Scenarios.} Performance in identifying correct fixes varies considerably with task difficulty: while \texttt{GPT-4} performs well at simpler fixes in a zero-shot setting, \texttt{CodeGemma} outperforms it on more complex variants, reaching 75.45\% accuracy. The introduction of in-context learning yields a dual effect—it significantly degrades performance for \texttt{GPT-4} and \texttt{CodeQwen}, yet consistently enhances it for open-source models like \texttt{Llama3.1-405B}, \texttt{CodeLlama-7B}, \texttt{CodeGemma}, and \texttt{Mistral}—highlighting the need for tailored adaptations to optimize in-context learning.

\noindent
\textbf{Predictive Scenarios.} The \texttt{Llama}-series models demonstrate robust capability in identifying target vulnerabilities, reflecting strong security-domain pre-training. In-context learning exerts mixed influences: models such as \texttt{GPT-4} and \texttt{Llama}-series display consistent improvements, whereas \texttt{CodeQwen1.5} and certain \texttt{Gemma} models face performance drops. These variations stress the significance of model-specific adaptation to capitalize on domain knowledge. Additionally, arbitrary prompts can degrade performance, mirroring the challenges noted in Goal-driven tasks. Hence, optimizing both prompt design and adaptation techniques remains pivotal for maximizing the effectiveness of LLMs in vulnerability analysis.

\begin{figure*}[!ht]
    \centering
    \begin{subfigure}{\textwidth}
        \centering
        \includegraphics[width=\textwidth]{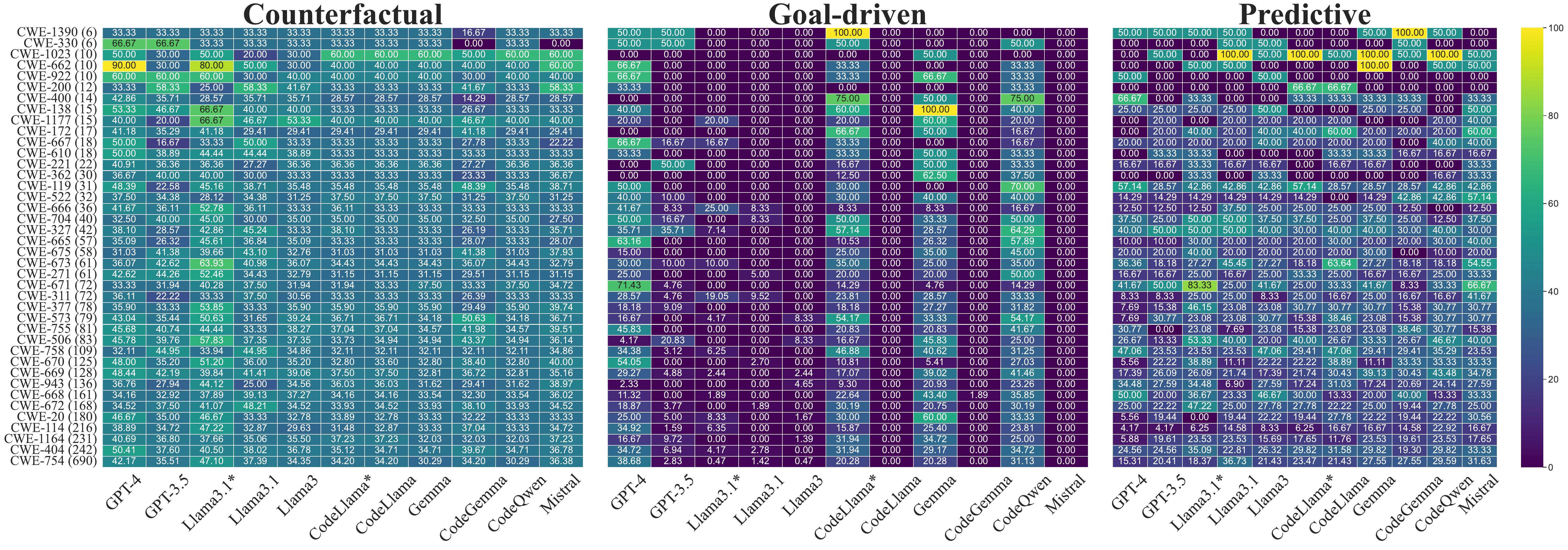}
        \caption{Zero-shot Inference}
        \label{fig:cwe_analysis_zero_shot}
    \end{subfigure}
    \begin{subfigure}{\textwidth}
        \centering
        \includegraphics[width=\textwidth]{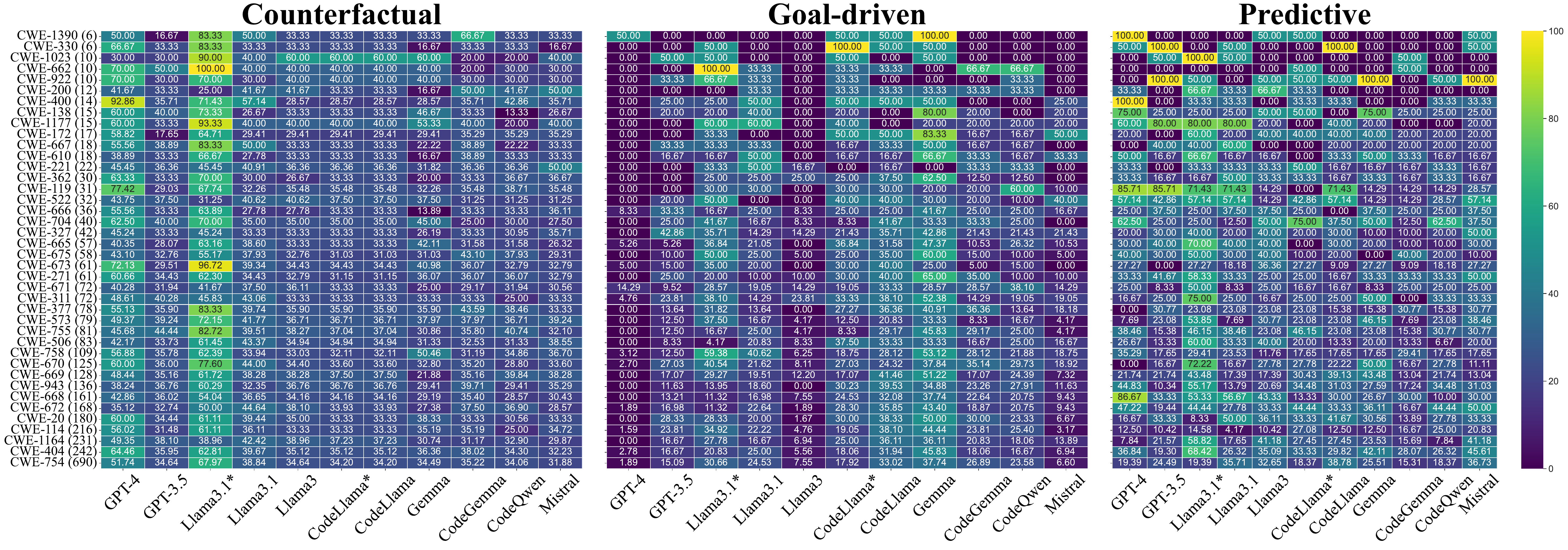}
        \caption{In-Context Learning Inference}
        \label{fig:cwe_analysis_incontext}
    \end{subfigure}
    
    \caption{Performance of LLMs across various question types for \textbf{class-level} CWEs under different inference modes. Models marked with $\star$ have larger parameter sizes (e.g., \texttt{Llama3.1}$^{\star}$ denotes \texttt{Llama3.1-405B}). The number in brackets beside each CWE indicates the total number of associated questions.}
    \label{fig:cwe_analysis}
\end{figure*}

\begin{table*}[ht]
\centering
\caption{Ablation study on \texttt{Llama3.1-8B} with varying inference temperatures. \textbf{Temp.} denotes the temperature setting.}
\resizebox{\textwidth}{!}{
\begin{tabular}{@{}l|cccccc|cccccccc|cccc@{}}
\toprule
\multirow{3}{*}{\textbf{Temp.}} & \multicolumn{6}{c|}{\textbf{Structure Reasoning}} & \multicolumn{8}{c|}{\textbf{Semantic Reasoning}} & \multicolumn{4}{c}{\textbf{Base Scenario}}\\
 & \multicolumn{2}{c}{\textbf{DataFlow}} & \multicolumn{2}{c}{\textbf{ControlFlow}} & \multicolumn{2}{c|}{\textbf{Average}} & \multicolumn{2}{c}{\textbf{Counterfactual}} & \multicolumn{2}{c}{\textbf{Goal-driven}} & \multicolumn{2}{c}{\textbf{Predictive}} & \multicolumn{2}{c|}{\textbf{Average}} & \multirow{2}{*}{\textit{\textbf{Unsafe}}} & \multirow{2}{*}{\textit{\textbf{Safe}}} & \multirow{2}{*}{\textbf{Average}} & \multirow{2}{*}{\textbf{FPR$^{\textbf{Safe}}$}}\\
 & \texttt{Zero} & \texttt{ICL} & \texttt{Zero} & \texttt{ICL} & \texttt{Zero} & \texttt{ICL} & \texttt{Zero} & \texttt{ICL} & \texttt{Zero} & \texttt{ICL} & \texttt{Zero} & \texttt{ICL} & \texttt{Zero} & \texttt{ICL} &  &  &  &  \\
\midrule
0.0 & 45.80 & \cellcolor{darkgreen!50}{\textcolor{white}{54.57}} & 54.94 & \cellcolor{darkgreen!50}{\textcolor{white}{52.64}} & 50.37 & \cellcolor{darkgreen!50}{\textcolor{white}{53.60}} & \cellcolor{darkgreen!50}{\textcolor{white}{36.87}} & \cellcolor{darkgreen!50}{\textcolor{white}{38.42}} & 1.12  & 23.04 & 24.90 & 27.54 & 20.96 & \cellcolor{lightgreencustom}{29.67} & 55.17 & \cellcolor{darkgreen!50}{\textcolor{white}{66.31}} & \cellcolor{darkgreen!50}{\textcolor{white}{60.74}} & \cellcolor{darkgreen!50}{\textcolor{white}{33.69}} \\
0.2 & \cellcolor{darkgreen!50}{\textcolor{white}{48.19}} & \cellcolor{lightgreencustom}{53.37}& 55.91 & \cellcolor{lightgreencustom}{51.75}& \cellcolor{lightgreencustom}{52.05}& \cellcolor{lightgreencustom}{52.56}& \cellcolor{lightgreencustom}{35.57}& \cellcolor{lightgreencustom}{35.78}& 12.94 & 25.02 & 25.73 & 27.54 & 24.75 & 29.45 & \cellcolor{lightgreencustom}{61.54} & \cellcolor{lightgreencustom}{53.58} & \cellcolor{lightgreencustom}{57.56} & 46.42 \\
0.4 & \cellcolor{lightgreencustom}{47.33}& 51.77 & \cellcolor{darkgreen!50}{\textcolor{white}{57.10}} & 51.08 & \cellcolor{darkgreen!50}{\textcolor{white}{52.22}} & 51.42 & 34.28 & 35.30 & 14.41 & 25.19 & 25.59 & \cellcolor{darkgreen!50}{\textcolor{white}{28.93}} & 24.76 & \cellcolor{darkgreen!50}{\textcolor{white}{29.81}} & \cellcolor{darkgreen!50}{\textcolor{white}{62.60}} & 47.21 & 54.91 & \cellcolor{lightgreencustom}{52.79} \\
0.6 & 44.65 & 46.34 & \cellcolor{lightgreencustom}{56.88}& 47.66 & 50.76 & 47.00 & 32.98 & 34.87 & 16.74 & 24.16 & 26.56 & \cellcolor{lightgreencustom}{28.51} & 25.43 & 29.18 & 60.48 & \cellcolor{lightgreencustom}{53.58} & 57.03 & 46.42 \\
0.8 & 43.00 & 44.94 & 52.71 & 47.58 & 47.86 & 46.26 & 33.06 & 34.90 & \cellcolor{lightgreencustom}{20.45} & \cellcolor{lightgreencustom}{25.63} & \cellcolor{lightgreencustom}{27.40} & 28.09 & \cellcolor{darkgreen!50}{\textcolor{white}{26.97}} & 29.54 & 57.82 & 49.07 & 53.45 & 50.93 \\
1.0 & 39.96 & 40.91 & 51.67 & 44.16 & 45.82 & 42.54 & 29.99 & 34.42 & \cellcolor{darkgreen!50}{\textcolor{white}{22.86}} & \cellcolor{darkgreen!50}{\textcolor{white}{27.26}} & \cellcolor{darkgreen!50}{\textcolor{white}{27.68}} & 27.12 & \cellcolor{lightgreencustom}{26.84} & 29.60 & 55.97 & 53.05 & 54.51 & 46.95 \\
\bottomrule
\end{tabular}
}
\label{tab:ablation_llama3_1_8B_highlight}
\end{table*}

\subsection{Behavior Distribution}
To further study the behavior of LLMs, we have detailed the distribution of choices made by LLMs during their evaluation, categorizing the options for each question into four types. In the DataFlow and ControlFlow-wise scenarios, the categories are: ``Direct C.'', indicating direct connections between code elements in the graph; ``Not C.'', denoting no significant connections; ``Indirect C.'', for indirect connections as described in our method section; and ``Unknown'', where the LLM is unable to respond. For the Counterfactual and Goal-driven scenarios, the responses are classified as ``Safe'', indicating that the current code variants are secure and do not trigger vulnerabilities; ``Bypass'', where the code neither triggers vulnerabilities nor executes the intended function; and ``Unsafe'', where the code variants do activate vulnerabilities. Lastly, in the Predictive scenario, ``Others'' refers to CWEs irrelevant to the question, ``Target'' denotes the CWEs specifically addressed in the question, and 'Bypass' represents variants that circumvent the targeted CWEs. Our key findings are as follows: 
\begin{enumerate} 
    \item \textbf{Pattern Matching Over Logic:} Our study reveals that LLMs in vulnerability analysis primarily rely on pattern matching based on pre-trained knowledge, rather than on logical analysis of the code. 
    \item \textbf{Need for Customized Approaches:} Effectively addressing vulnerabilities with LLMs requires more than general in-context learning; it demands advanced prompt engineering or scenario-specific fine-tuning. 
    \item \textbf{Need for Domain Adaption:} LLMs often struggle to differentiate between various CWEs, underscoring a critical need for domain adaptation to enhance their effectiveness in vulnerability analysis.
\end{enumerate}

Specifically, as illustrated in Figure~\ref{fig:choice_dirtribution}, in the Dataflow-wise and Controlflow-wise scenarios, we observe that LLMs like \texttt{GPT-3.5}, \texttt{CodeLlama}, \texttt{CodeQwem} and \texttt{Gemma} predominantly choose options classified as ``Direct C.'' and ``Indirect C.'' This suggests these LLMs assume that all code elements in a given piece of code are interconnected. These findings align with our analysis in Section~\ref{sec:effects_of_difficulty}, confirming that such LLMs lack rigorous logical analysis capabilities to understand contextual relationships in code. Instead, their vulnerability analysis tends to rely heavily on pattern matching, a conclusion supported by a substantial body of existing research~\cite{mirchandani2023large,weber2024large,mirzadeh2024gsm,xie2024memorization}. Introducing in-context learning slightly mitigates this tendency, though a substantial gap remains. In the Counterfactual scenario, some LLMs cannot differentiate cases of ``Bypass'' code, struggling to correctly determine the code's execution path. This so-called runtime reasoning ability~\cite{chen2024reasoning} is crucial for vulnerability analysis, particularly for runtime-dependent vulnerabilities like Buffer Errors and Input Validation. Additionally, many LLMs show a distribution of responses heavily skewed toward ``unsafe'', implying the code contains vulnerabilities. Notably, this happens even when our generator places vulnerable code along a dead path in the scenario. This phenomenon further supports the hypothesis that LLMs rely on pattern matching rather than genuine runtime reasoning.

In the more complex Goal-driven scenario, most LLMs tend to select ``Unknown.'' For instance, \texttt{GPT-4} and \texttt{LLama}-series models, although they perform relatively well in other scenarios, frequently select ``Unknown'' option in this scenario. This indicates that current LLMs have difficulty applying parameterized knowledge to resolve vulnerabilities effectively. Unlike in other use cases, addressing vulnerabilities requires advanced prompt engineering or fine-tuning customized for specific scenarios. Lastly, in the Predictive scenario, most LLMs distribute their choices evenly across all available options. Incorrect classifications such as ``Other,'' ``Bypass,'' and ``Unknown'' dominate their responses. This distribution suggests that most LLMs:
% \begin{enumerate}
    % \item 
    i) cannot reliably determine whether a vulnerability would be triggered in the current scenario. Simple pattern matching is insufficient for this level of judgment;
    % \item 
    and ii) struggle to distinguish fundamental differences between CWEs.
% \end{enumerate}
This highlights the need for domain adaptation and specialized approaches. Without these, LLMs are not well-suited for direct application to vulnerability analysis.
\begin{figure}[ht]
    \centering
    \includegraphics[width=0.49\textwidth]{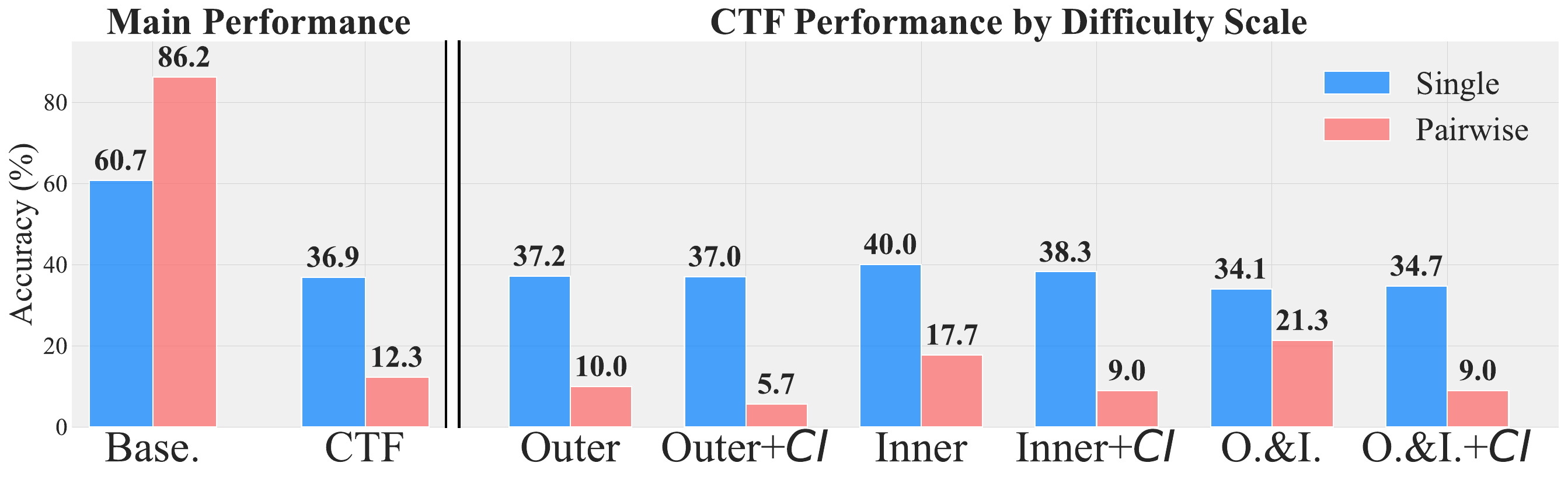}
    \caption{Pairwise performance analysis. \textbf{Base} and \textbf{CTF} indicate base and Counterfactual scenarios performance, respectively. $+$\textit{CI} denotes code variants with control-flow injection. “Single” means one question per thread, while “Pairwise” means questions with similar context are combined for inference.}
    \label{fig:pairwise}
\end{figure}
\begin{table*}[ht]
\centering
\caption{Comparison of LLM performance on benchmark generated based on \textbf{PrimeVul} Dataset~\cite{ding2024vulnerability}.}
\resizebox{\textwidth}{!}{
\begin{tabular}{@{}l|cccccc|cccccccc|cccc}
\toprule
\multirow{3}{*}{\textbf{Models}} & \multicolumn{6}{c|}{\textbf{Structure Reasoning}} & \multicolumn{8}{c|}{\textbf{Semantic Reasoning}} & \multicolumn{4}{c}{\textbf{Base Scenario}}\\

 & \multicolumn{2}{c}{\textbf{DataFlow}} (601) & \multicolumn{2}{c}{\textbf{ControlFlow}} (419) & \multicolumn{2}{c|}{\textbf{Average}} & \multicolumn{2}{c}{\textbf{Counterfactual}} (658) & \multicolumn{2}{c}{\textbf{Goal-driven}} (189) & \multicolumn{2}{c}{\textbf{Predictive}} (205) & \multicolumn{2}{c|}{\textbf{Average}} & \multirow{2}{*}{\textit{\textbf{Unsafe}} (64)} & \multirow{2}{*}{\textit{\textbf{Safe}} (64)}  & \multirow{2}{*}{\textbf{Average}} & \multirow{2}{*}{\textbf{FPR$^{\textbf{Safe}}$}}\\
 
 & \texttt{Zero} & \texttt{ICL} & \texttt{Zero} & \texttt{ICL} & \texttt{Zero} & \texttt{ICL} & \texttt{Zero} & \texttt{ICL} & \texttt{Zero} & \texttt{ICL} & \texttt{Zero} & \texttt{ICL} & \texttt{Zero} & \texttt{ICL} \\

\midrule
\multicolumn{19}{c}{\cellcolor{gray!10}$>$\textbf{15B Param. Models}} \\
\midrule
\texttt{GPT-4-turbo} &\cellcolor{lightgreencustom}{51.91} & \cellcolor{lightgreencustom}{55.75}\textcolor{deepskyblue}{$\uparrow$} & \cellcolor{darkgreen!50}{\textcolor{white}{73.75}} & \cellcolor{darkgreen!50}{\textcolor{white}{76.27}}\textcolor{deepskyblue}{$\uparrow$} & \cellcolor{darkgreen!50}{\textcolor{white}{62.83}} &  \cellcolor{darkgreen!50}{\textcolor{white}{66.01}}\textcolor{deepskyblue}{$\uparrow$} & 44.83 & 48.98\textcolor{deepskyblue}{$\uparrow$} & 13.23 & 21.40\textcolor{deepskyblue}{$\uparrow$} & 14.15 & 17.17\textcolor{deepskyblue}{$\uparrow$} & 24.07 & 29.18\textcolor{deepskyblue}{$\uparrow$} & 17.19 & 81.25 & 49.22 & \cellcolor{lightgreencustom}{18.75}\\
    
\texttt{GPT-3.5-turbo} &35.27 & 37.11\textcolor{deepskyblue}{$\uparrow$} & 45.35 & 46.15\textcolor{deepskyblue}{$\uparrow$} & 40.31 & 42.63\textcolor{deepskyblue}{$\uparrow$} & \cellcolor{lightgreencustom}{45.74} & \cellcolor{lightgreencustom}{49.35}\textcolor{deepskyblue}{$\uparrow$} & 28.57 & 29.10\textcolor{deepskyblue}{$\uparrow$} & 21.95 & 22.08\textcolor{deepskyblue}{$\uparrow$} &  \cellcolor{darkgreen!50}{\textcolor{white}{32.09}} & \cellcolor{lightgreencustom}{33.51}\textcolor{deepskyblue}{$\uparrow$} & 32.81 & 62.50 & 47.66 & 37.50\\

\texttt{Llama3.1-405B}$^{\star}$ &\cellcolor{darkgreen!50}{\textcolor{white}{54.41}} & \cellcolor{darkgreen!50}{\textcolor{white}{55.07}}\textcolor{deepskyblue}{$\uparrow$} & \cellcolor{lightgreencustom}{70.88} & \cellcolor{lightgreencustom}{71.84}\textcolor{deepskyblue}{$\uparrow$} & \cellcolor{lightgreencustom}{62.64} & \cellcolor{lightgreencustom}{63.46}\textcolor{deepskyblue}{$\uparrow$} & \cellcolor{darkgreen!50}{\textcolor{white}{50.76}} & \cellcolor{darkgreen!50}{\textcolor{white}{57.75}}\textcolor{deepskyblue}{$\uparrow$} & 24.87 & \cellcolor{darkgreen!50}{\textcolor{white}{31.22}}\textcolor{deepskyblue}{$\uparrow$} & 10.24 & 15.12\textcolor{deepskyblue}{$\uparrow$} & 28.62 & \cellcolor{darkgreen!50}{\textcolor{white}{34.70}}\textcolor{deepskyblue}{$\uparrow$} & 85.94 & 15.62 & 50.78 & 84.38\\

\midrule
\multicolumn{19}{c}{\cellcolor{gray!10}$<$\textbf{15B Param. Models}} \\
\midrule 
\texttt{Llama3.1-8B}$^{\star}$ &42.43 & 47.75\textcolor{deepskyblue}{$\uparrow$} & 53.70 & 57.76\textcolor{deepskyblue}{$\uparrow$} & 48.06 & 52.76\textcolor{deepskyblue}{$\uparrow$} & 28.72 & 33.89\textcolor{deepskyblue}{$\uparrow$} & 17.46 & 12.70\textcolor{crimsonred}{$\downarrow$} & \cellcolor{darkgreen!50}{\textcolor{white}{27.32}} & \cellcolor{lightgreencustom}{28.29}\textcolor{deepskyblue}{$\uparrow$} & 24.5 & 24.96\textcolor{deepskyblue}{$\uparrow$} & 40.62 & 68.75 & \cellcolor{darkgreen!50}{\textcolor{white}{54.69}} & 31.25\\

\texttt{Llama3-8B}$^{\star}$ &41.60 & 44.76\textcolor{deepskyblue}{$\uparrow$} & 49.16 & 55.13\textcolor{deepskyblue}{$\uparrow$} & 45.38 & 49.94\textcolor{deepskyblue}{$\uparrow$} & 35.87 & 36.47\textcolor{deepskyblue}{$\uparrow$} & 4.76 & 11.64\textcolor{deepskyblue}{$\uparrow$} & 12.20 & 25.37\textcolor{deepskyblue}{$\uparrow$} & 17.61 & 24.49\textcolor{deepskyblue}{$\uparrow$} & 21.88 & 81.25 & \cellcolor{lightgreencustom}{51.56} & \cellcolor{darkgreen!50}{\textcolor{white}{18.75}}\\

\texttt{Gemma-7B}$^{\star}$ &35.11 & 34.28\textcolor{crimsonred}{$\downarrow$} & 21.96 & 21.96\textcolor{crimsonred}{$\downarrow$} & 28.54 & 28.12\textcolor{crimsonred}{$\downarrow$} & 29.48 & 27.05\textcolor{crimsonred}{$\downarrow$} & \cellcolor{lightgreencustom}{32.80} & \cellcolor{lightgreencustom}{30.16}\textcolor{crimsonred}{$\downarrow$} & 23.41 & \cellcolor{darkgreen!50}{\textcolor{white}{30.73}}\textcolor{deepskyblue}{$\uparrow$} & 28.56 & 29.31\textcolor{deepskyblue}{$\uparrow$} & 98.44 & 0.00 & 49.22 & 100.00\\

\texttt{CodeQwen1.5-7B}$^{\star}$ &33.61 & 31.45\textcolor{crimsonred}{$\downarrow$} & 26.49 & 33.89\textcolor{deepskyblue}{$\uparrow$} & 30.05 & 32.67\textcolor{deepskyblue}{$\uparrow$} & 34.50 & 35.11\textcolor{deepskyblue}{$\uparrow$} & \cellcolor{darkgreen!50}{\textcolor{white}{35.45}} & 26.46\textcolor{crimsonred}{$\downarrow$} & \cellcolor{lightgreencustom}{24.88} & 26.34\textcolor{deepskyblue}{$\uparrow$} & \cellcolor{lightgreencustom}{31.61} & 29.30\textcolor{crimsonred}{$\downarrow$} & 71.88 & 0.00 & 35.94 & 100.00\\

\texttt{Mixtral-7B}$^{\star}$ &33.61 & 41.93\textcolor{deepskyblue}{$\uparrow$} & 26.49 & 42.96\textcolor{deepskyblue}{$\uparrow$} & 30.05 & 42.44\textcolor{deepskyblue}{$\uparrow$} & 29.64 & 29.33\textcolor{crimsonred}{$\downarrow$} & 1.59 & 4.23\textcolor{deepskyblue}{$\uparrow$} & 20.98 & 26.34\textcolor{deepskyblue}{$\uparrow$} & 17.40 & 19.97\textcolor{deepskyblue}{$\uparrow$} & 0.00 & 0.00 & 0.00 & 100.00\\
\toprule
\end{tabular}}
\label{tab:main_res_primevul}
\end{table*}

\subsection{Pairwise Performance Evaluation}
\textbf{Pairwise Performance Evaluation} assesses pairs of samples that share the same functionality but differ in vulnerability. In our single Q\&A setting, each question was presented in a single prompt per thread. By contrast, in the pairwise setting, each safe\&unsafe pair was presented together in one prompt per thread. Both experiments were conducted independently. Inspired by PrimeVul~\cite{ding2024vulnerability}, we use these pairs to examine how effectively LLMs handle syntactically similar yet semantically distinct scenarios within the same context. In the ``Single'' evaluation, each question $Q$ is presented independently, producing $A = f(\{Q\})$, where $A$ is the answer and $f$ the LLM. In the ``Pairwise'' evaluation, a set of questions $\{Q_1, \ldots, Q_N\}$ is presented simultaneously, yielding $\{A_1, \ldots, A_N\} = f(\{Q_1, \ldots, Q_N\})$. We measure performance in the ``Single'' setting as 
$\tfrac{\#\text{(Correct Answers)}}{\#\text{(Total Questions)}}$ and in the ``Pairwise'' setting as
$\tfrac{\#\text{(Correct Pairs)}}{\#\text{(Total Pairs)}}$, where a pair is correct only if \textbf{all} its answers are correct.

We restricted pairwise evaluations to the \emph{Base} and \emph{Counterfactual} scenarios using \texttt{Llama3.1-8B}, excluding Goal-driven and Predictive scenarios because they do not produce semantically distinct safe\&unsafe label pairs required for meaningful pairwise comparison. As shown in Figure~\ref{fig:pairwise}, performance in the Base scenario increases from 60.7\% to 86.2\% under pairwise evaluation, but drops from 36.9\% to 12.3\% in the Counterfactual scenario, consistently across all difficulty levels. We also examined the consistency score in the pairwise setting, $\text{Cons}_{\text{CTF}}^{\text{Pairwise}}$, defined as:
\[
\text{Cons}_{\text{CTF}}^{\text{Pairwise}} 
= \frac{\sum_{i=1}^{N_{\text{CTF}}^{\text{Pairwise}}} \mathbb{I}\Bigl(C_{\text{CTF}^{\text{Pairwise}}}^i = 1 \,\wedge\, C_{\text{base}^{\text{Pairwise}}}^i = 1\Bigr)}
       {N_{\text{CTF}}^{\text{Pairwise}}}.
\]
This measure extends the consistency score introduced in Eq.~\ref{eq:CTF1} for the ``single'' setting. Under single evaluation, it is 24.3\%, whereas under pairwise evaluation, it drops to 15.3\%. A likely explanation involves the Juliet dataset, which is widely available and may have been part of the model’s pre-training data. Since Juliet often places safe and vulnerable code side by side, \texttt{Llama3.1-8B} effectively recognizes these patterns in the Base Scenario. However, the Counterfactual Scenario relies on our generator to generate semantically perturbed variants that deviate from the model's training data, resulting in a marked performance decline. This outcome further indicates that LLMs tend to rely more on pattern matching than on deeper logical reasoning.

\subsection{Effects of CWEs}  
The CWE hierarchy comprises pillars, classes, bases, and variants~\cite{martin2008common}. Figure~\ref{fig:cwe_analysis} shows our evaluation of various LLMs (zero-shot and in-context) mapped from Juliet’s CWEs to their \textbf{class-levels}. We observe notable performance variations tied to the LLMs’ pre-training domains. For example, \texttt{Gemma} and \texttt{CodeLlama} excel in Goal-driven scenarios like CWE-20 and CWE-1390, indicating a need for targeted enhancements in other CWEs. Comparing zero-shot and in-context heatmaps reveals that in-context learning typically improves understanding of specific CWE classes; however, models like \texttt{GPT-4} may underperform in certain Goal-driven cases. Overall, these findings underscore the importance of tailoring prompt designs to each CWE for effective vulnerability analysis.

\subsection{Effects of Temperature} 
We conducted an ablation study on inference temperature (ranging from 0.0 to 1.0 in 0.2 increments) using both zero-shot and in-context learning with \texttt{Llama3.1-8B}. As shown in Table~\ref{tab:ablation_llama3_1_8B_highlight}, lower temperatures (e.g., 0.0) produce deterministic outputs that perform well in tasks requiring consistency, such as structural reasoning, base scenarios, and Counterfactual questions. Conversely, for more challenging tasks like Goal-driven and Predictive scenarios, higher temperatures yield better performance by fostering greater output diversity, which allows the model to explore varied reasoning paths and generate more creative answers.

\subsection{Generalization of SV-TrustEval-C}  
We evaluated our generator and benchmark on the \textbf{PrimeVul} dataset, which provides higher label accuracy and broader CWE coverage than other sources. We manually validated samples from each CWE category and selected one verified \emph{safe} vs. \emph{unsafe} pair per category. Note that we skipped compilation validation, as PrimeVul lacks a direct compilation source—a step required to match the precision of our Juliet experiments. This process generated 64 \emph{safe} and 64 \emph{unsafe} base scenario samples, along with 601 DataFlow-based, 419 ControlFlow-based, 658 Counterfactual, 189 Goal-driven, and 205 Predictive questions, all of which are included in our GitHub repository\footnote{\url{https://github.com/Jackline97/SV-TrustEval-C}}.

Table~\ref{tab:main_res_primevul} shows trends similar to those observed with Juliet. Larger models (e.g., \texttt{GPT-4}, \texttt{GPT3.5}, \texttt{Llama3.1-405B}) perform well in structural tasks, while most models struggle with semantic tasks, often scoring below 50\%. High false positive rates persist, with some models misclassifying safe code as unsafe. In-context learning generally boosts performance—especially for models with limited code pre-training (e.g., \texttt{CodeQwen}, \texttt{CodeGemma}, \texttt{Mixtral})—although gains vary. Overall, these findings confirm that \textbf{SV-TRUSTEVAL-C} generalizes well to datasets with broader CWE coverage, though it is advisable to use datasets that support full quality verification. (i.e., label accuracy and compilation validation).

\section{Conclusion}
We present \textsc{SV-TrustEval-C}, our initial benchmark for evaluating LLMs’ vulnerability reasoning in C. Our study shows that most LLMs, including \texttt{GPT-4}, rely more on pattern matching than logical analysis, struggle with semantic reasoning, and perform inconsistently across code variants. They are also easily misled by bypass attempts, indicating limited structural understanding and domain-specific knowledge. These findings highlight the need to improve LLMs’ code analysis. Future work will expand \textsc{SV-TrustEval-C} to additional programming languages and more complex scenarios.

\bibliographystyle{IEEEtran}
\bibliography{citation}

\appendices

\begin{figure*}[t]
    \centering
\includegraphics[width=0.95\textwidth]{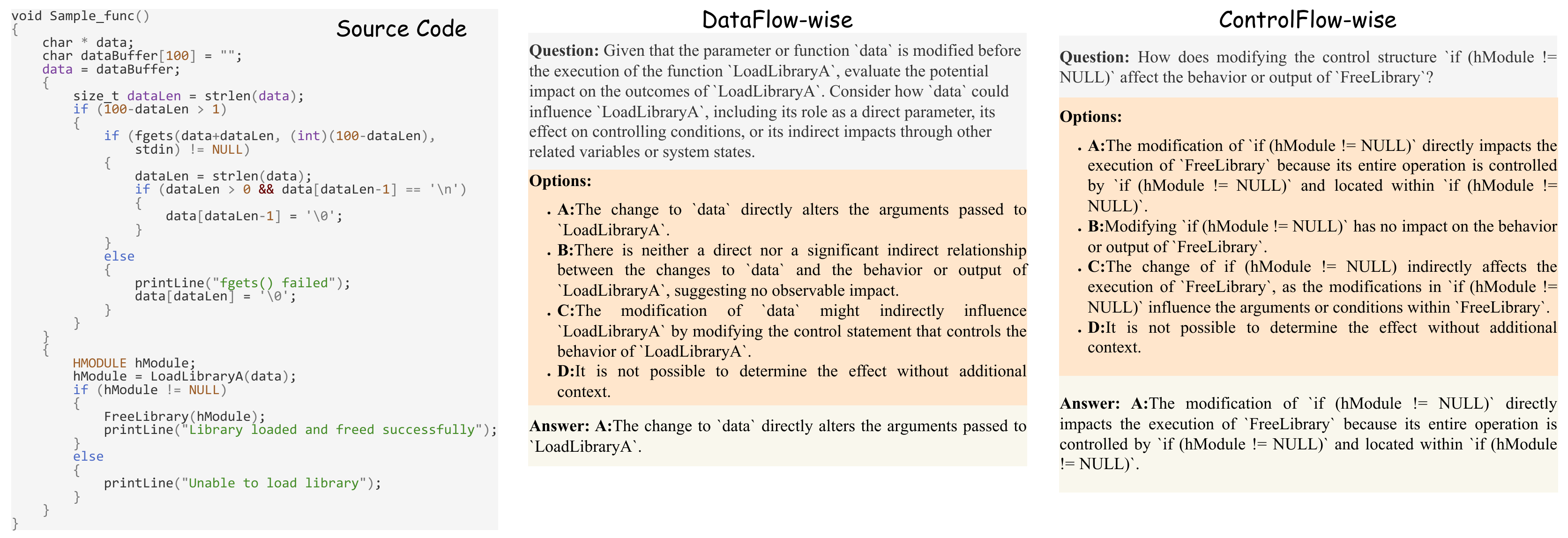}
    \caption{Examples of Structure Reasoning Questions.}
    \label{fig:SR_sample}
\end{figure*}

\section{Question Examples}\label{sec:sample}

\begin{figure}[ht]
    \centering
\includegraphics[width=0.45\textwidth]{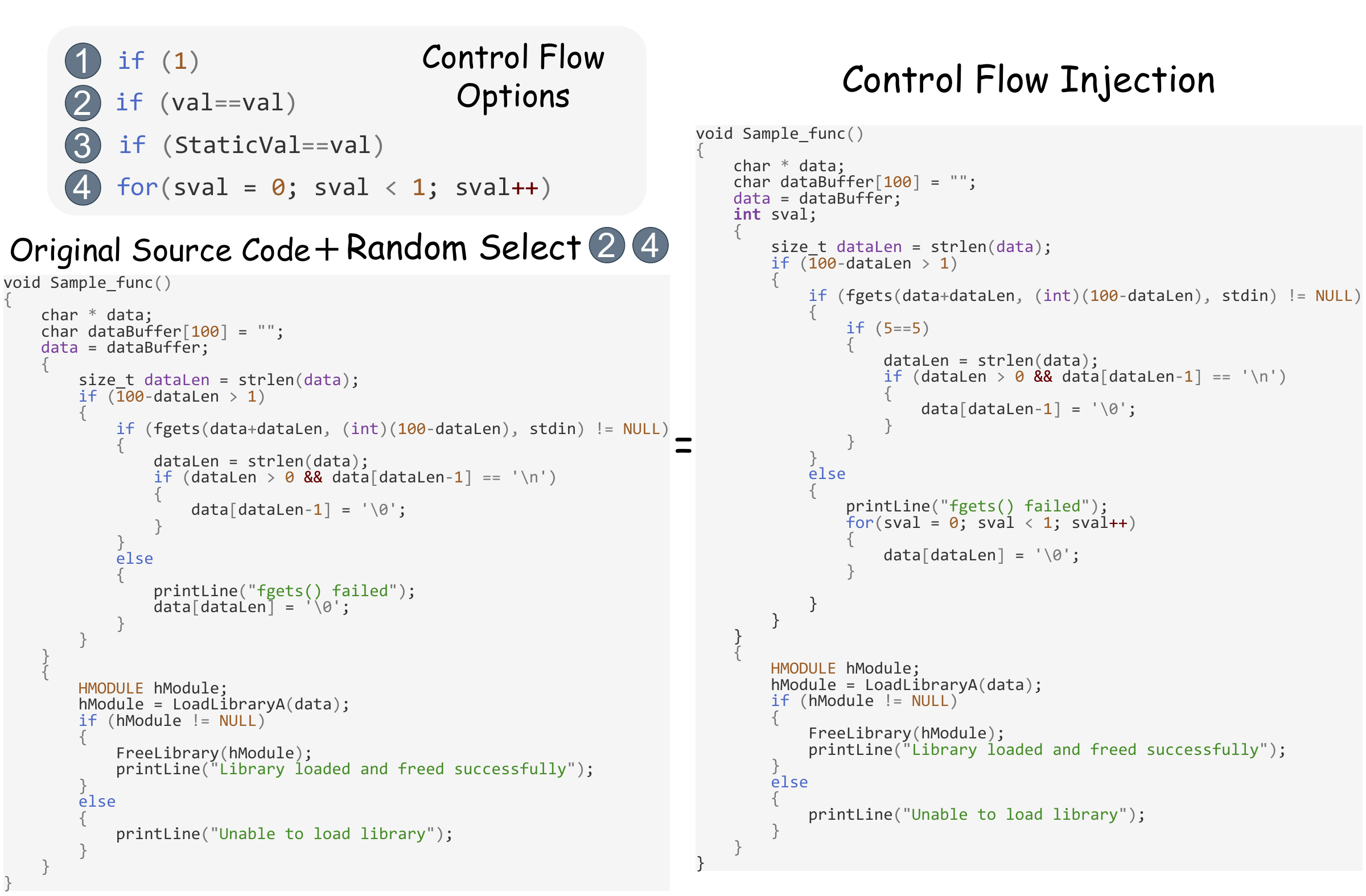}
    \caption{Examples of Control Flow Injection.}
    \label{fig:contol_inject}
\end{figure}

\begin{figure}[ht]
    \centering
    \includegraphics[width=0.45\textwidth]{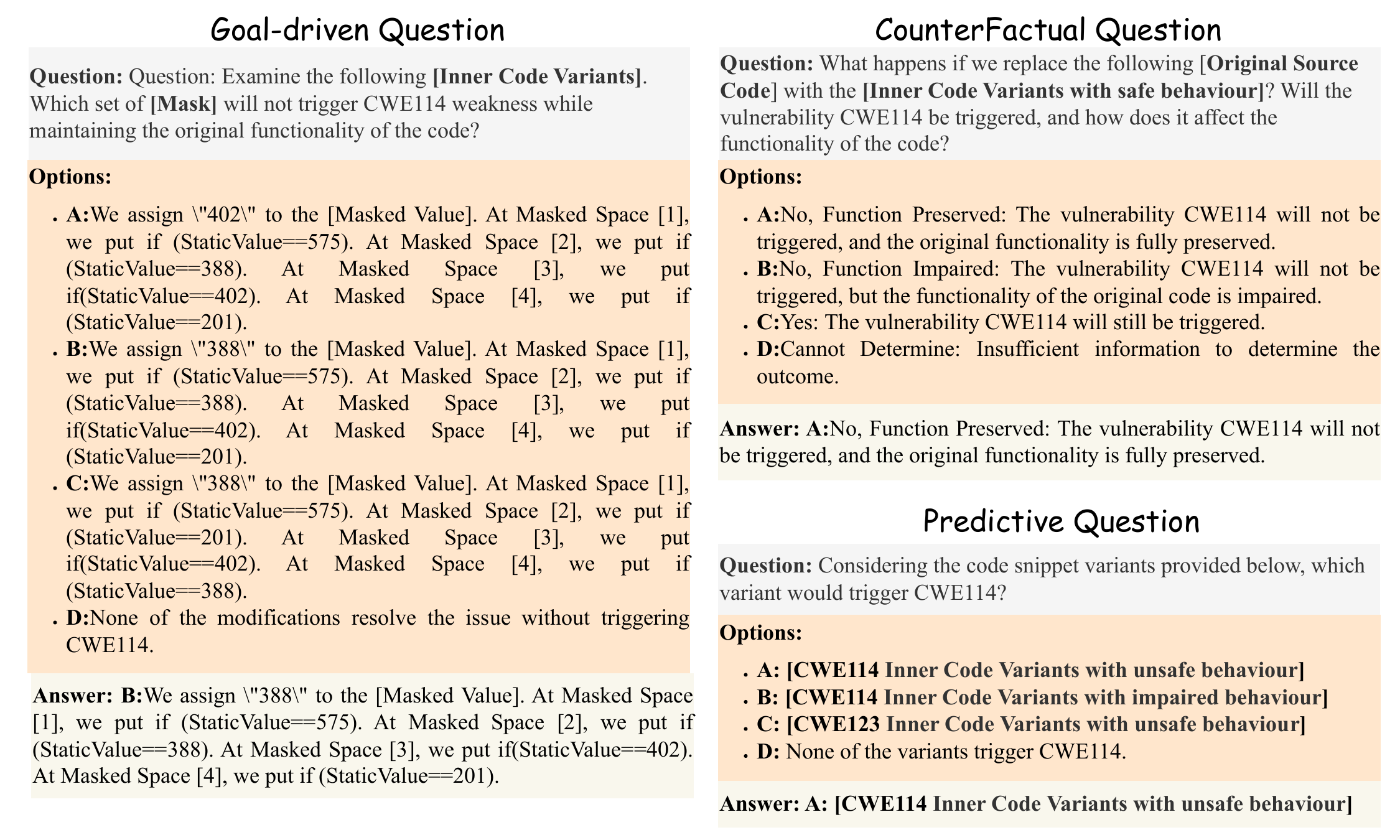}
    \caption{Examples of Semantic Reasoning Questions.}
    \label{fig:semantic_sample}
\end{figure}

Our SV-TrustEval-C Benchmark is demonstrated through example questions. Figure~\ref{fig:SR_sample} shows how structural reasoning questions are derived from the original code, while Figure~\ref{fig:contol_inject} presents our Variants Generator, which adds control flow branches to create Outer, Inner, and Outer\&Inner variants with behaviors (safe, unsafe, impaired) that underpin the semantic reasoning questions in Figure~\ref{fig:semantic_sample}.

\subsection{Impaired Code Sample}

% \lstset{
%   basicstyle=\small\ttfamily,
%   columns=flexible,
%   breaklines=true,
%   postbreak=\mbox{\textcolor{red}{$\hookrightarrow$}\space},
%   aboveskip=0pt,
%   belowskip=0pt
% }

% \begin{lstlisting}[language=Python, label={lst:example}]
% {
%     int socketDescriptor;
%     int connectionStatus = 0;
%     char networkBuffer[256];
%     for (socketDescriptor = 0; socketDescriptor < 5; socketDescriptor++) {
%         if (connectionStatus == 0)
%             networkBuffer[socketDescriptor] = 'A';
%         else
%             networkBuffer[socketDescriptor] = '0';
%     }
%     printLine("Successful! Problem Resolved!");
% }
% \end{lstlisting}

\begin{figure}[t]
    \centering
    \includegraphics[width=0.4\textwidth]{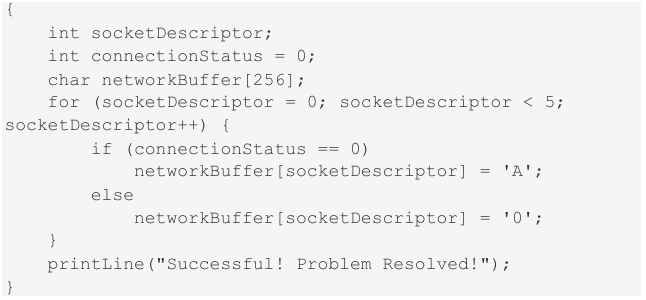}
    \caption{Overview of Impaired Code}
    \label{fig:Impairedcode}
\end{figure}

We introduced the snippet in Figure~\ref{fig:Impairedcode} as impaired code $C^{\text{impaired}}$, simulating false success to test if LLMs grasp the original code's vulnerabilities.

\section{Error Case Analysis}
\begin{figure}[h]
    \centering
    \includegraphics[width=0.49\textwidth]{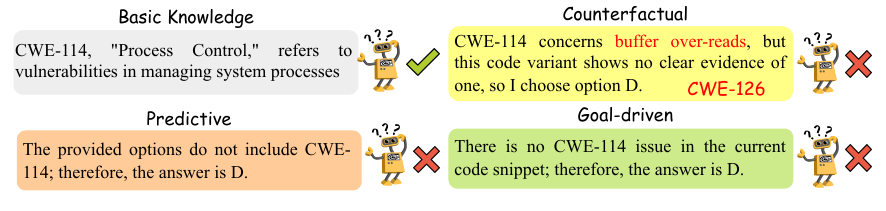}
    \caption{Error cases analysis of \texttt{GPT-4}}
    \label{fig:error_case}
\end{figure}
% \subsubsection{Effects of CWEs with ICL}\label{sec:cwe_effect}
% By comparing the heatmaps for zero-shot and in-context learning across various semantic reasoning scenarios, as illustrated in the Figure~\ref{fig:cwe_analysis_incontext}, we observe that in-context learning generally enhances or activates certain LLMs' knowledge of specific CWE classes, thereby improving their performance in most cases. Additionally, different prompt designs have distinct effects on different CWEs. Although models like \texttt{GPT-4} show decreased performance in Goal-driven scenarios, the overall results demonstrate that applying specific LLMs to particular CWEs in vulnerability analysis requires customized prompt designs to optimize their effectiveness.

Our analysis shows that 33\% of zero-shot and 21\% of in-context semantic errors result in "Do not know" responses due to inconsistent vulnerability interpretation. For example, as Figure~\ref{fig:error_case} illustrates, GPT-4 initially identifies CWE-114 correctly but later misclassifies it or responds "I don't know," indicating an overreliance on pattern matching over deep semantic reasoning.
\newpage

\section{Meta-Review}

The following meta-review was prepared by the program committee for the 2025
IEEE Symposium on Security and Privacy (S\&P) as part of the review process as
detailed in the call for papers.

\subsection{Summary}
This paper introduces SV-TrustEval-C, a novel benchmark designed to evaluate the reasoning capabilities of Large Language Models (LLMs) in detecting source code vulnerabilities. The benchmark focuses on two key dimensions: structural reasoning, which assesses how well models identify relationships between code elements under varying data and control flow complexities, and semantic reasoning, which evaluates their logical consistency when code is altered both structurally and semantically. The authors evaluate 11 LLMs on this benchmark, revealing that current models rely more on pattern matching than on robust logical reasoning for vulnerability analysis.

\subsection{Scientific Contributions}
\begin{itemize}
\item Creates a New Tool to Enable Future Science: The SV-TrustEval-C benchmark provides a systematic way to evaluate LLMs' reasoning capabilities in vulnerability analysis, enabling future research in this area.
\item Provides a Valuable Step Forward in an Established Field: The paper advances the field of LLM-based vulnerability detection by introducing a novel approach that leverages multiple-choice questions to assess reasoning, moving beyond traditional prompt engineering methods.
\end{itemize}

\subsection{Reasons for Acceptance}
\begin{enumerate}
\item The SV-TrustEval-C benchmark is a significant contribution to the field, as it allows for the systematic evaluation of LLMs' reasoning capabilities in vulnerability analysis. The benchmark is designed to be extensible, enabling future researchers to apply it to other datasets and codebases.

\item The paper addresses a critical gap in the evaluation of LLMs for vulnerability detection by focusing on both structural and semantic reasoning. The findings highlight the limitations of current models, providing valuable insights for future improvements in LLM-based vulnerability analysis.

\item The paper includes a detailed evaluation of 11 LLMs, offering a broad perspective on the current state of LLM performance in vulnerability detection. The results underscore the need for further research into enhancing the logical reasoning capabilities of LLMs.
\end{enumerate}

\subsection{Noteworthy Concerns}
\begin{enumerate}
\item Bias in the Benchmark: While the authors have addressed concerns about bias by removing keywords like "good" and "bad" from the Juliet Test Suite, there is still a potential for bias due to the reliance on a single dataset. The authors have taken steps to mitigate this by extending their evaluation to the PrimeVul dataset, but further validation on real-world codebases would strengthen the benchmark's generalizability.

\item Generalization Beyond Juliet Dataset: The benchmark's reliance on the Juliet Test Suite raises questions about its applicability to real-world vulnerabilities. The authors have demonstrated extensibility by applying their approach to the PrimeVul dataset, but further validation on open-source vulnerability datasets (e.g., CVE-based datasets) would enhance the benchmark's credibility.
\end{enumerate}

\section{Response to the Meta-Review}
We appreciate the reviewers’ insightful feedback and the shepherd’s guidance in improving our paper. While we acknowledge the value of extending our benchmark to additional real-world codebases, we must also maintain its quality in terms of label accuracy and code compilation validation. Our use of the Juliet and PrimeVul datasets is informed by these considerations: Juliet includes a compilable check environment for our generator and its artificially generated nature ensures label accuracy, whereas PrimeVul incorporates real-world codebases with rigorous label validation. Although further exploration of additional real-world datasets would likely reinforce our conclusions, the extensive effort required for label verification and code-compilation checks goes beyond our current scope. We plan to address this in future work.

\end{document}